\documentclass{article}

\usepackage[eandd,final]{neurips_2026}

\usepackage[utf8]{inputenc} % allow utf-8 input
\usepackage[T1]{fontenc}    % use 8-bit T1 fonts
\usepackage{hyperref}       % hyperlinks
\usepackage{url}            % simple URL typesetting
\usepackage{booktabs}       % professional-quality tables
\usepackage{amsfonts}       % blackboard math symbols
\usepackage{nicefrac}       % compact symbols for 1/2, etc.
\usepackage{microtype}      % microtypography
\usepackage{xcolor}         % colors
\usepackage{amssymb} 
\usepackage{graphicx}
\usepackage{amsmath} 
\usepackage{adjustbox}
\usepackage[most]{tcolorbox}
\usepackage{caption}
\usepackage{pifont}
\usepackage{enumitem}
\usepackage{multirow}
\newtcolorbox{insightbox}{
  colback=gray!5,
  colframe=gray!40,
  boxrule=0.4pt,
  arc=2pt,
  left=4pt,
  right=4pt,
  top=3pt,
  bottom=3pt
}

\newtcolorbox{insightline}{
  blank,
  boxrule=0pt,
  boxsep=0pt,
  left=6pt,
  right=0pt,
  top=2pt,
  bottom=2pt,
  borderline west={1.2pt}{0pt}{gray!40!white},
  colback=gray!3!white,
}

\newtcolorbox{insightblue}{
  colback=blue!3!white,
  colframe=blue!50!black,
  boxrule=0.3pt,
  arc=2pt,
  boxsep=0pt,
  left=8pt, right=8pt, top=6pt, bottom=6pt,
  title={\textbf{Insight}},
  coltitle=blue!50!black,
  fonttitle=\bfseries
}
\usepackage{xcolor}
\usepackage{colortbl}
\usepackage{booktabs}
\usepackage{enumitem}
\usepackage{titlesec}

\definecolor{tablegray}{RGB}{248,248,248}

\definecolor{tab1}{RGB}{198,228,255}
\definecolor{tab2}{RGB}{222,240,255}

\definecolor{tab3}{RGB}{208,238,208}
\definecolor{tab4}{RGB}{226,246,226}

\definecolor{tab5}{RGB}{255,252,220}

\definecolor{tab6}{RGB}{240,235,255}

\definecolor{vdbTeal}{HTML}{7FB3F0}
\definecolor{vdbCoral}{HTML}{FFB6D5}
\definecolor{vdbHeader}{HTML}{EDF2F4}
\definecolor{vdbGreen}{HTML}{B0B0DA}

\usepackage[dvipsnames]{xcolor}
\usepackage{tabularx}

\hypersetup{
    colorlinks=true,
    linkcolor=Maroon,
    citecolor=NavyBlue,
    urlcolor=OliveGreen
}

\title{VecDBLens: An Evaluatology-Based Framework for Vector Database Diagnosis}
\author{%
  \parbox{0.95\textwidth}{%
    \centering
    Yutong Zhou\textsuperscript{1,2},
    Guoxin Kang\textsuperscript{1},
    Wenxin Zhou\textsuperscript{1,2}\thanks{Wenxin Zhou and Haoyu He contributed equally and share third authorship.},
    Haoyu He\textsuperscript{3}\footnotemark[1],
    Yuedong Zhu\textsuperscript{4} \\[0.3em]
    Lei Wang\textsuperscript{1},
    Wanling Gao\textsuperscript{1},
    Fanda Fan\textsuperscript{1},
    Jianfeng Zhan\textsuperscript{1} \\[0.6em]
    \textsuperscript{1}Institute of Computing Technology,
    Chinese Academy of Sciences\\
    \textsuperscript{2}University of Chinese Academy of Sciences\\
    \textsuperscript{3}The Chinese University of Hong Kong, Shenzhen\\
    \textsuperscript{4}Universiti Sains Malaysia%
  }%
}

\begin{document}

\maketitle

\begin{abstract}
Vector databases are increasingly deployed as retrieval pipelines in which representation, indexing, and query-time execution jointly determine accuracy, throughput, and cost.
Existing vector database evaluations primarily focus on indexing while treating embeddings as fixed inputs and ignoring the role of query-time execution, making it difficult to decide whether a bottleneck should be addressed by changing the embedding model, the index, or the query-processing functions.
To address the above limitation, we introduce \textbf{VecDBLens}, an Evaluatology-based framework that treats the function under diagnosis as an \emph{evaluated object} and all remaining dataset, workload, parameter, and runtime factors as matched \emph{evaluation conditions}. VecDBLens attributes performance differences to individual evaluated objects through comparisons under matched evaluation conditions, summarizing their performance and stability across multiple datasets and workloads. It further analyzes interactions among objects using analysis of variance (ANOVA), thereby turning vector database benchmarking from leaderboard ranking into a reproducible diagnostic procedure.
VecDBLens reveals two key insights. First, retrieval performance is fundamentally upper-bounded by embedding quality, revealing that many apparent indexing failures are therefore representation-side limitations. Second, stronger index parameterization is not always beneficial: overly aggressive configurations can reshape neighborhood topology and cause recall regression. These results challenge the common practice of evaluating vector databases primarily through indexing and show that real-world limitations arise from constraints across retrieval functions rather than any single function alone.  VecDBLens provides a principled basis for identifying the true performance drivers of vector databases and for guiding their design in practical applications. The source code is available from \url{https://github.com/zhouyutong20/VecDBLens}.
\end{abstract}

%broad evaluation

\section{Introduction}

\textbf{Background.} Vector databases have become core infrastructure for modern retrieval applications, including semantic search, recommendation, retrieval-augmented generation, and multimodal retrieval~\cite{qdrant, wang2021milvus, johnson2019billion, 10.1007/s00778-024-00864-x, gao2023retrieval}.  Their observed performance, however, is not determined by an ANN index in isolation~\cite{malkov2018efficient, jegou2011product, hnswlib, ootomo2023cagra, 8681160, wang2021comprehensivesurveyexperimentalcomparison}. A deployed vector retrieval pipeline first maps raw objects into a representation space, then organizes vectors into an index, and finally executes query-time search together with runtime constraints.

\textbf{Limitation and Motivation.} Existing benchmarks have produced valuable progress in comparing ANN algorithms, typically under fixed embeddings and a narrow range of configurations~\cite{patel2024acorn, jiang2025digra, rangepq}. This design is appropriate for measuring approximation loss inside a given vector space, yet it is insufficient for diagnosing end-to-end performance in real applications. The same low recall can arise from a weak \textbf{embedding} that places relevant objects far apart, an \textbf{index} that fails to retain the embedding-defined neighborhood, or a \textbf{query-processing} strategy that becomes inefficient under filtering or other constraints.

This paper asks a diagnostic question: \emph{when a vector retrieval pipeline fails to meet a target Recall--QPS trade-off, which object is actually responsible?} Answering this question requires evaluating objects under matched evaluation conditions rather than ranking whole databases under a fixed benchmark setting.

\begin{figure}[t]
    \centering
    \includegraphics[width=1\linewidth]{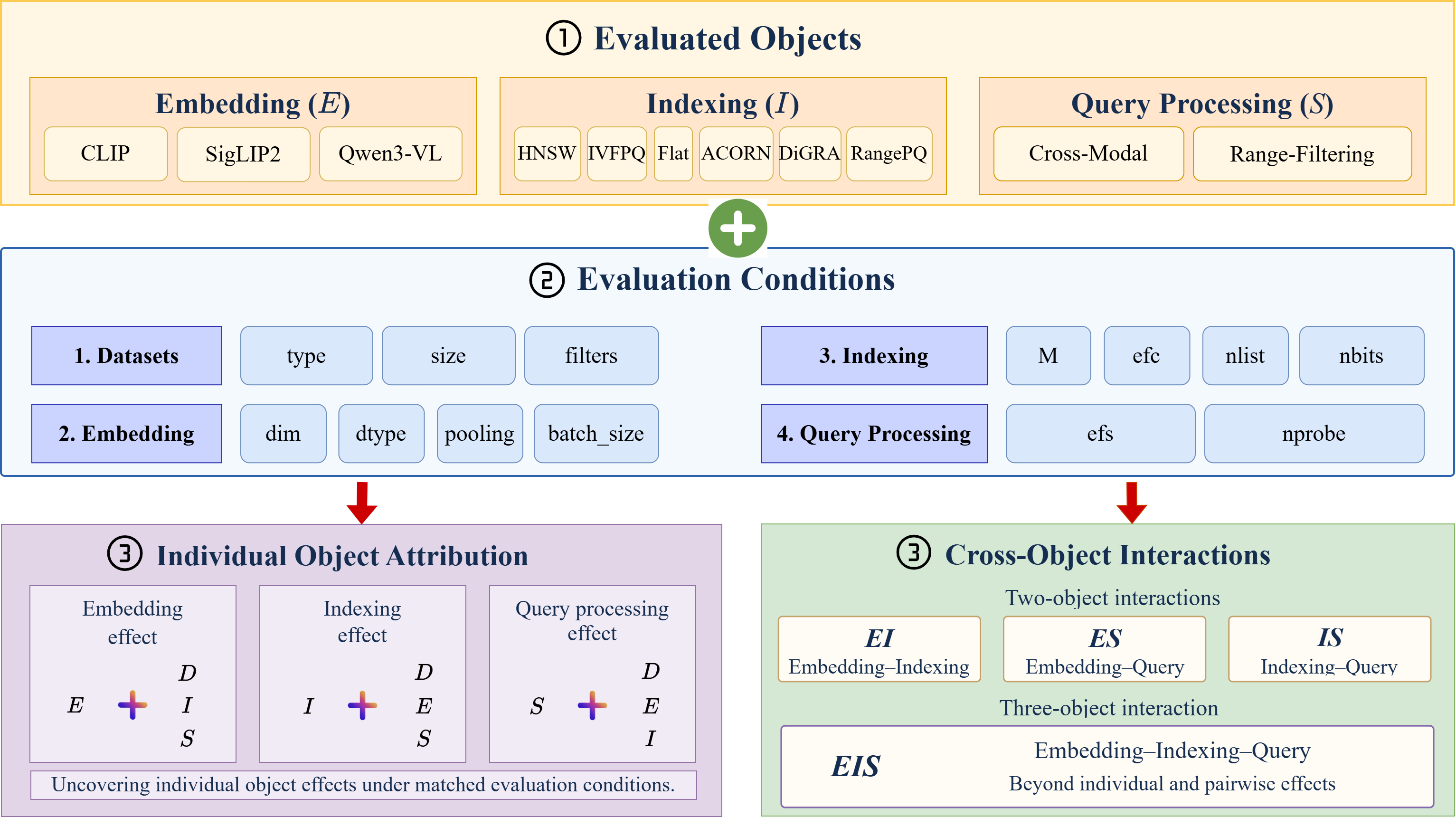}
    \caption{
Overview of VecDBLens. 
VecDBLens first decomposes a vector database into three evaluated objects: embedding, indexing, and query processing, and defines an evaluation condition space covering datasets and object configurations. 
It then attributes performance differences to individual evaluated objects through comparisons under matched evaluation conditions and analyzes interactions among objects using ANOVA, enabling systematic diagnosis of performance variation and benchmark insights.}
\label{fig:framework}
\end{figure}

We identify two evaluation biases that prevent such a diagnosis: (1) \textbf{\emph{index-centric attribution.}} 
Existing benchmarks often define ground truth using distance-based nearest neighbors, which may deviate from the natural relevance relationships in real-world cross-modal retrieval. 
As a result, retrieval failures caused by weak or misaligned embeddings may be incorrectly attributed to index structures. 
(2) \textbf{\emph{narrow configuration space.}}
Although existing benchmarks vary configurations such as graph connectivity, probing budget, or search depth, they often treat embeddings, workloads, and filtering regimes as fixed. 
This can favor vector databases whose optimal operating regimes happen to match the benchmark setting, while underestimating systems that are effective under different evaluation conditions.

\textbf{Contribution.} To address these issues, we propose \textbf{VecDBLens}, shown in Figure~\ref{fig:framework}. VecDBLens decomposes the retrieval pipeline into \textbf{embedding}, \textbf{indexing}, and \textbf{query processing} functions. For each diagnostic question, the function under diagnosis is the \emph{evaluated object} (EO), while all remaining dataset, workload, parameter, and hardware factors form the \emph{evaluation conditions} (ECs). By comparing EOs only over shared or stratified ECs, VecDBLens estimates individual object effects across matched evaluation conditions instead of conflating them with dataset or workload variation. VecDBLens separates the \textbf{individual object effects across matched evaluation conditions} from the extrinsic variation introduced by datasets, workloads, filter selectivity, parameter settings, and runtime environments. We summarize our contributions as follows: 

First, we identify two key biases in existing vector database benchmarks: \textbf{index-centric attribution} and \textbf{narrow configuration space}. We formulate a diagnostic framework for vector retrieval pipelines that distinguishes evaluated objects from their evaluation conditions and supports both \textbf{attribution to individual objects} and \textbf{analysis of interactions among objects}.

Second, we introduce a reproducible diagnostic protocol that attributes performance differences to individual evaluated objects \textbf{under matched evaluation conditions} and analyzes interactions among objects \textbf{using analysis of variance (ANOVA)}. This protocol goes beyond leaderboard comparison by explaining whether performance variation is driven by embedding, indexing, or query-time execution.

Finally, we instantiate \textbf{VecDBLens} on two complementary workloads: multimodal text-to-image retrieval and filtered ANN search. The experiments show that \textbf{embedding quality} imposes an effective upper bound on retrieval performance, and \textbf{stronger index parameterization} is not monotonically beneficial.

\section{Related Work}

\subsection{Vector Databases, Applications, and Benchmarks} 
The architecture of vector databases~\cite{10.1007/s00778-024-00864-x, wang2021milvus, 10.14778/3554821.3554843, 10.14778/3685800.3685805} has fundamentally evolved from basic Approximate Nearest Neighbor (ANN) search~\cite{10.1145/276698.276876,jegou2011product,malkov2018efficient, johnson2019billion, wang2021comprehensivesurveyexperimentalcomparison} to complex databases designed for real-world application demands, such as cross-modal retrieval and filtered ANN search~\cite{10.1145/3639324,10.5555/3692070.3692567,10.1145/3698814,10.14778/3717755.3717770,10.1145/3769843}. Early developments predominantly focused on ANN index performance. For instance, \textbf{Faiss}~\cite{johnson2019billion} integrated various indexing algorithms like \textbf{HNSW}~\cite{malkov2018efficient} and \textbf{IVFPQ}~\cite{jegou2011product}, while libraries such as \textbf{hnswlib}~\cite{hnswlib} pioneered the standalone, highly optimized implementation of the \textbf{HNSW} algorithm.

However, isolated ANN indexes struggle with the inherent complexities of real-world scenarios—specifically, overcoming the embedding blind spots in cross-modal data understanding and addressing the overhead of scalar data processing~\cite{10.1145/3769843,10.1145/3543507.3583552}. To resolve these limitations, modern vector databases have emerged, integrating embedding generation as a retrieval function. \textbf{Vector databases} like \emph{Milvus}~\cite{wang2021milvus} and \emph{Qdrant}~\cite{qdrant}, for example, natively integrate \textbf{embedding models} such as \emph{CLIP}~\cite{radford2021clip}, SigLIP2~\cite{tschannen2025siglip2} and Qwen3-VL~\cite{bai2025qwen3vl} to facilitate seamless cross-modal search. Similarly, to efficiently handle filtered ANN search, specialized solutions including \emph{RangePQ} ~\cite{rangepq}, \emph{ACORN}~\cite{patel2024acorn}, and \emph{DiGRA}~\cite{jiang2025digra} have been introduced. These approaches optimize query execution through techniques such as range space partitioning and dynamic adjustment of graph connectivity based on filtering selectivity.

Diverging from existing benchmarks~\cite{benchmark,BigVectorBench,big-ann-benchmarks} that primarily focus on end-to-end rankings or approximation curves, this paper diagnoses the retrieval pipeline as a set of interacting functions: \textbf{\emph{embedding, indexing, and query processing}}. The goal is to provide a reproducible diagnostic procedure that identifies whether the limiting factor is the representation space, the index approximation, or the query-time execution.

\subsection{Evaluation Methodology}
Traditional ANN benchmarks primarily assess index recall and efficiency within a given vector space~\cite{benchmark,big-ann-benchmarks}. Iceberg~\cite{10.1145/3786690} extends evaluation to downstream tasks, including image classification, face recognition, text retrieval, and recommendation, examining embedding loss, metric misuse, and distribution sensitivity. These studies motivate further diagnosis of how retrieval functions affect application-level performance.

Firstly, they predominantly rely on synthetic workloads to reveal how different index structures, parameter settings, and execution backends affect recall, latency, and system cost~\cite{benchmark}. A critical flaw of these synthetic workloads is that the nearest neighbors are purely derived from  \emph{mathematical distance metrics}, rather than emerging naturally from real-world applications. Our experiments demonstrate that in real-world cross-modal retrieval scenarios, in the worst cases, a mere  \textbf{2.04\%}\footnote{On VisualNews, using SigLIP2 embeddings, we perform exhaustive Euclidean search with Faiss CPU \texttt{IndexFlatL2} over 1,259,732 candidate images for 20,000 caption queries, with both image and text embeddings L2-normalized. A query is counted as a hit if its nearest image is the paired image specified by the dataset. There are 408 hits, yielding Recall@1 = $408/20{,}000 = 2.04\%$.} of the nearest neighbors calculated via distance metrics correspond to the true semantic nearest neighbors.

Secondly, because existing benchmarks strictly focus on ANN indexing, most existing evaluations treat embeddings as  \emph{fixed} input vectors and define exact nearest neighbors within the given vector space as the ground truth by  \emph{brute-force search}~\cite{8681160,benchmark,big-ann-benchmarks,wang2021comprehensivesurveyexperimentalcomparison}. As a result, they can explain how much performance is lost during {approximate search}, but they usually cannot further  \emph{attribute end-to-end retrieval performance} to different pipeline functions.

To overcome these limitations, we introduce VecDBLens, a framework that builds on the pipeline functions decomposition and attribution methods of Evaluatology~\cite{zhan2024evaluatology, zhan2025evaluatology}, which has been applied to the evaluation of CPUs~\cite{wang2025achieving, wang2026poster}, large language models~\cite{tianbeyond}, and time-series forecasting models~\cite{wang2026combinationts}.
By systematically sampling across an extensive configuration space of datasets and function parameters to construct rigorous evaluation conditions, VecDBLens quantifies both average effects and stability of distinct functions,  making it possible to distinguish consistent function behavior across conditions from condition-specific wins.

\section{The VecDBLens Framework}

Existing benchmarks primarily evaluate indexes under fixed embeddings and a narrow range of configurations. Distance-based neighbors can differ from application-level relevance, while different retrieval configurations can perform best under different workloads. These limitations make it difficult to determine whether a retrieval bottleneck originates from embedding, indexing, or query processing. VecDBLens addresses this diagnostic question by attributing performance differences to individual evaluated objects under matched evaluation conditions and quantifying interactions among objects using ANOVA~\cite{sobol1990sensitivity}.

\subsection{Evaluation System Modeling}
\textbf{Problem statement.} VecDBLens decomposes the retrieval pipeline into embedding, indexing, and query processing. For a dataset instance $d$, let $\operatorname{corpus}(d)$ denote its retrieval corpus and $\mathcal{Q}_d$ its query set. Given a query $q\in\mathcal{Q}_d$, the pipeline is
\begin{equation}
\mathcal{R}_{q}^{e,a,\beta,\gamma} = S_{\gamma}\left(I_{a,\beta}\left(E_e(\operatorname{corpus}(d))\right), E_e(q)\right).
\label{eq:vecdb_pipeline}
\end{equation}
The embedding function $E_e$ maps raw objects into dense vectors $E_e(\operatorname{corpus}(d))$ and $E_e(q)$~\cite{bai2025qwen3vl, radford2021clip}. The indexing function $I_{a,\beta}$ constructs a searchable structure $I_{a,\beta}(E_e(\operatorname{corpus}(d)))$ using algorithm $a$ and construction parameters $\beta$, such as HNSW~\cite{malkov2018efficient}, IVFPQ~\cite{jegou2011product}, or CAGRA~\cite{ootomo2023cagra}. The query-processing function $S_\gamma$ searches the index under query-time parameters $\gamma$, trading off recall and efficiency. Flat search provides an exact-search reference for the selected embedding and similarity function.

\textbf{Evaluated objects and evaluation conditions.}  
Guided by Evaluatology~\cite{zhan2025evaluatology,wang2026poster,wang2025achieving,tianbeyond,wang2026combinationts}, we construct a \textbf{Self-Contained System (SCS)} for the systematic evaluation of vector databases. The SCS comprises evaluated objects (EOs) and evaluation conditions (ECs).

The evaluated object, denoted by $\mathrm{eo}$, is the model, algorithm, or configuration under diagnosis. An evaluation condition instance, denoted by $\mathrm{ec}$, specifies the values of the remaining factors for that evaluation. A factor selected as the EO in one analysis can therefore be held fixed as part of $\mathrm{ec}$ in another. As shown in Figure~\ref{fig:framework}, when the \textbf{embedding model} is the EO, the dataset, index configuration, search settings, and runtime environment are ECs. When the \textbf{index algorithm} is the EO, the embedding becomes an EC. The full notation and condition-space definition are provided in Appendix~\ref{appendix:notation}.

\subsection{Individual Object Attribution and Cross-object Interactions}

\label{sec:distributional_metrics}

Based on the SCS model, VecDBLens first systematically evaluates an object's performance across matched evaluation conditions, then examines how its effect depends on other evaluated objects.

\textbf{Individual object attribution.} 
VecDBLens constructs the EC space from the Cartesian product of the selected factor values~\cite{enderton1977elements}, retaining compatible configurations. Each point in this space represents one evaluation condition instance $\mathrm{ec}$. For individual object attribution, VecDBLens varies $\mathrm{eo}$ while holding $\mathrm{ec}$ fixed. For indexing, for example, ACORN~\cite{patel2024acorn} and RangePQ~\cite{rangepq} are compared under identical embedded data, queries, search settings, and runtime resources. The evaluated differences in recall, throughput, or cost quantify the performance effect attributable to the EO choice under the matched evaluation condition. VecDBLens aggregates these differences with equal weighting across dataset or workload strata, and reports mean performance and within-stratum variance to characterize effectiveness and stability.

\textbf{Cross-object interactions.} Vector embedding, index construction, and query processing interact. VecDBLens distinguishes these objects for attribution without assuming independence. All evaluation results come from complete end-to-end retrieval pipelines with compatible configurations of all three objects. When one object is selected as the evaluated object (EO), the other two form part of the matched evaluation conditions (ECs).
VecDBLens examines these interactions by testing whether the performance difference attributed to one object changes with the configuration of another object. For example, repeating the construction comparison at different search budgets reveals whether more extensive traversal reduces the recall difference between constructed indexes.

Within each index family, VecDBLens evaluates complete matched grids of embedding, construction, and query-processing configurations, fixing the dataset, candidate pool, queries, and runtime environment. Functional ANOVA~\cite{sobol1990sensitivity} decomposes Retention@10 and $\log_2(\mathrm{QPS})$ into main effects and two- and three-factor interactions. Main effects summarize each factor's average effect across the other configurations; interaction terms capture departures from their additive combination. Their variance shares quantify interaction strength, while conditional comparisons explain its direction. Formal attribution metrics, grid weights, and interaction definitions are provided in Appendix~\ref{sec:metric_definitions}.

\section{Experiments}
\label{sec:experiments}

This section instantiates VecDBLens on two representative scenarios:
(1) \textbf{Multimodal retrieval scenario (In Sec.~\ref{subsec:exp_embedding_ceiling})}. We diagnose how much retrieval quality is determined before indexing and how much of this fixed-representation potential is retained by ANN backends.
(2) \textbf{Filtered ANN search scenario (In Sec.~\ref{subsec:exp_filtered_ann})}. We diagnose whether stronger index parameterization consistently improves performance, and how workload conditions, such as filter ratio and dimensionality, reshape the useful regime of different methods.

\textbf{Statistical protocol details.}
For each EO, VecDBLens summarizes performance across evaluated ECs.
We report \textbf{mean performance}, \textbf{stability} measured by average within-stratum variance, \textbf{bootstrap confidence intervals}, and \textbf{observed maximum and minimum values}.
Direct comparisons between EOs use matched evaluation conditions where available.
For throughput, we report \textbf{geometric mean QPS} because throughput varies by orders of magnitude.

\subsection{Evaluation Setup}
\label{subsec:exp_setup}

We instantiate the \textbf{VecDBLens} Evaluated Object (EO) and Evaluation Condition (EC) framework across two representative scenarios: multimodal retrieval and filtered ANN search. These scenarios provide complementary perspectives on ANN evaluation: cross-modal retrieval examines semantic alignment between queries and retrieved objects, while filtered ANN examines interactions between scalar constraints and ANN indexes. Both have practical relevance to multimedia search, multimodal RAG, and vector queries with scalar constraints. Runtime settings and parameter ranges are provided in Appendix~\ref{appendix:configs}.

\textbf{Multimodal Retrieval.}
The configuration space is defined as $\Theta_{\mathrm{mm}} = \mathcal{E} \times \mathcal{A} \times \mathcal{B} \times \Gamma$, representing embeddings, ANN backends, construction parameters, and search parameters. 
For embedding attribution, the EO is an embedding model $e \in \{\mathrm{CLIP}, \mathrm{SigLIP2}, \mathrm{Qwen3\text{-}VL}\}$~\cite{radford2021clip,tschannen2025siglip2,bai2025qwen3vl}, evaluated against a Flat exact-search EC to expose the representational upper bound.

For ANN attribution, the EO is a backend $a \in \{\mathrm{Faiss\ HNSW},\allowbreak
\mathrm{Faiss\ IVFPQ},\allowbreak
\mathrm{cuVS\ IVFPQ},\allowbreak
\mathrm{cuVS\ CAGRA}\}$~\cite{johnson2019billion,jegou2011product,ootomo2023cagra} along with its parameters $(\beta,\gamma)$, while the selected embedding becomes an EC. Retention is measured relative to the corresponding Flat reference.
The data context spans text-to-image retrieval tasks across Flickr30K (test5k/full), MSCOCO (test5k/full), and VisualNews~\cite{young2014flickr30k,lin2014mscoco,liu2021visualnews}.

\textbf{Filtered ANN Search.}
The configuration space is $\Theta_{\mathrm{filter}} = \mathcal{A}_{f} \times \mathcal{B}_{f} \times \Gamma_{f} \times \mathcal{T}$, representing algorithms, construction parameters, search parameters, and runtime settings.
The evaluated EOs include Flat, RangePQ, DiGRA, ACORN-1, and ACORN-28~\cite{rangepq,jiang2025digra,patel2024acorn}. Algorithm-level comparisons use matched thread settings; ACORN is tested under single- and multi-threaded settings with matched construction and search configurations to assess parallel execution effects.
For parameter attribution, the EO is a controlled parameter change, such as increasing $M$ or \texttt{efSearch}, with the remaining factors fixed as ECs. Parameters linked by a configuration rule are changed jointly.
The experiments use million-scale workloads derived from the widely used SIFT1M and GIST1M benchmarks~\cite{jegou2011product,jegou2010sift,oliva2001modeling}, with dimensionalities dim $\in \{128,960\}$ and scalar attributes for controlled range filtering. Unlike cross-modal retrieval, the main challenge is index traversal under scalar constraints. Filtering preserves the distance function, so exact neighbors within the eligible set provide ground truth for approximation accuracy.
We evaluate filter ratios $r \in \{0.01\%, 0.1\%, 1\%, 10\%, 50\%\}$ and use Flat exact pre-filtering as the reference for each workload condition.

To extend VecDBLens to a broader range of languages and representation formats, we additionally evaluate Chinese text retrieval on MIRACL and multi-vector retrieval on SciFact and FiQA, with experimental settings and results provided in Appendix~\ref{app:additional_scenarios}.

\subsection{Experiment 1: Embedding Quality and ANN Retention}
\label{subsec:exp_embedding_ceiling}

This experiment applies a two-level attribution. First, we treat the embedding model instance $e$ as the EO and the dataset-pool setting $d$ as the EC, using Flat search to measure retrieval quality under exact search. Second, holding the embedding $e$ and dataset-pool setting $d$ fixed, we treat the ANN backend and its configuration $(a,\beta,\gamma)$ as the EO and compare its recall with the corresponding Flat reference, separating representation quality from changes introduced by indexing and query processing.

\subsubsection{Embedding-induced upper bound.}
For a fixed embedding model instance $e$ and dataset instance $d$, Flat exact search yields the best attainable retrieval quality:
\begin{equation}
U(e,d,k) = R_k(e,\mathrm{Flat},\varnothing,\varnothing,d).
\label{eq:flat_upper_bound}
\end{equation}
Rather than replacing ground-truth labels, $U(e,d,k)$ quantifies the intrinsic geometric alignment of relevant objects in the representation space. If an embedding space maps relevant pairs far apart, downstream ANN indexes cannot systematically recover this lost relevance.

As shown in Table~\ref{tab:flat_upper_bound}, Qwen3-VL achieves the highest mean Flat Recall@10 across five conditions (0.6988), outperforming CLIP (0.5230) and SigLIP2 (0.2225). This indicates that a large portion of retrieval variation is predetermined before indexing. Many observed failures are fundamentally representation-side limitations.

To distinguish embedding-model effects from differences in output dimensionality and assess the uncertainty of matched-condition comparisons, we conduct matched-dimension experiments and cluster-aware significance tests (Appendix~\ref{app:embedding_controls}).

\begin{table*}[t]
\centering

\caption{Flat exact-search Recall@10 across embedding models and dataset-pool conditions. Mean and sample variance are computed across five conditions. The 95\% CI for the mean is estimated by condition-level bootstrap.}

\label{tab:flat_upper_bound}
\resizebox{\linewidth}{!}{%
\begin{tabular}{lccccc ccc}
\toprule
\textbf{Embedding}
& \textbf{Flickr30K}
& \textbf{Flickr30K}
& \textbf{MSCOCO}
& \textbf{MSCOCO}
& \textbf{VisualNews}
& \textbf{Mean}
& \textbf{Variance}
& \textbf{95\% CI}
\\
& \textbf{test5k}
& \textbf{full-pool}
& \textbf{test5k}
& \textbf{full-pool}
& 
& 
& 
& 
\\
\midrule
% Qwen3-VL
% & \textcolor{red}{\underline{0.9732}}
% & \textcolor{red}{\underline{0.7562}}
% & \textcolor{red}{\underline{0.8960}}
% & \textcolor{red}{\underline{0.5998}}
% & \textcolor{red}{\underline{0.2687}}
% & \textcolor{red}{\underline{0.6988}}
% & \textcolor{blue}{0.0781}
% & [0.4637, 0.8989]
% \\
% CLIP
% & 0.9008
% & 0.4946
% & 0.6727
% & 0.2973
% & 0.2496
% & 0.5230
% & 0.0729
% & [0.3177, 0.7345]
% \\
% SigLIP2
% & \textcolor{blue}{0.5982}
% & \textcolor{blue}{0.2246}
% & \textcolor{blue}{0.1685}
% & \textcolor{blue}{0.0488}
% & \textcolor{blue}{0.0724}
% & \textcolor{blue}{0.2225}
% & \textcolor{red}{\underline{0.0492}}
% & [0.0822, 0.4183]
Qwen3-VL
&\cellcolor{vdbTeal!59} \underline{0.9732}
&\cellcolor{vdbTeal!47} \underline{0.7562}
&\cellcolor{vdbTeal!54} \underline{0.8960}
&\cellcolor{vdbTeal!38} \underline{0.5998}
&\cellcolor{vdbTeal!20} \underline{0.2687}
&\cellcolor{vdbTeal!43} \underline{0.6988}
& 0.0781
& [0.4637, 0.8989]
\\
CLIP
&\cellcolor{vdbTeal!55} 0.9008
&\cellcolor{vdbTeal!32} 0.4946
&\cellcolor{vdbTeal!42} 0.6727
&\cellcolor{vdbTeal!21} 0.2973
&\cellcolor{vdbTeal!19} 0.2496
&\cellcolor{vdbTeal!34} 0.5230
& 0.0729
& [0.3177, 0.7345]
\\
SigLIP2
&\cellcolor{vdbTeal!38} 0.5982
&\cellcolor{vdbTeal!17} 0.2246
&\cellcolor{vdbTeal!14} 0.1685
&\cellcolor{vdbTeal!8} 0.0488
&\cellcolor{vdbTeal!9} 0.0724
&\cellcolor{vdbTeal!17} 0.2225
& \underline{0.0492}
& [0.0822, 0.4183]
\\
\bottomrule
\end{tabular}
}
\end{table*}

% \begin{insightbox}
% \finding{Insight 1. Embedding quality establishes the strict upper bound for retrieval performance.}
% \end{insightbox}
\begin{insightline}
\textbf{Insight 1. \emph{Embedding quality establishes the strict upper bound for retrieval performance.}}
\end{insightline}

\subsubsection{Candidate-pool scaling.}
We examine whether exact-search retrieval quality is robust to distractor scaling by expanding the candidate pool from \texttt{test5k} to \texttt{full-pool}. We define Pool-Scaling Retention as $\operatorname{PSR}(e,d_{\mathrm{test5k}},d_{\mathrm{full}},k) = U(e,d_{\mathrm{full}},k) / U(e,d_{\mathrm{test5k}},k)$, which measures the retention of exact-search recall under larger pools. We macro-average PSR across Flickr30K and MSCOCO.

Figure~\ref{fig:exp1_recall10} shows that Qwen3-VL retains a macro-averaged 72.3\% of its Recall@10 after expansion, compared with CLIP (49.6\%) and SigLIP2 (33.2\%). Stronger embeddings achieve higher fixed-pool recall and exhibit greater robustness to candidate-pool expansion in these experiments. Detailed numerical results are reported in Appendix~\ref{sec:detailed_results}.

\begin{insightline}
\textbf{Insight 2. \emph{High-quality embeddings preserve more of their exact-search recall under candidate-pool expansion.}}
\end{insightline}

\subsubsection{ANN retention relative to Flat.}

% zadd2 合并的表>U<
\begin{figure}[t]
    \centering
    \includegraphics[width=0.9\textwidth]{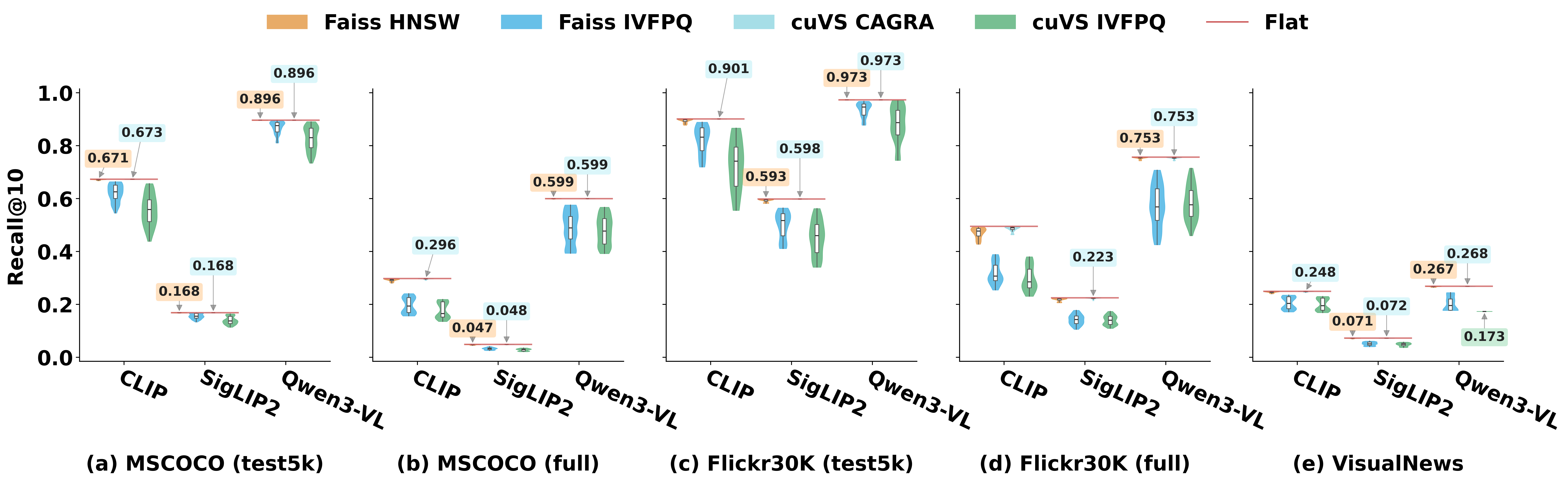}
    \caption{Recall@10 distributions on MSCOCO, Flickr30K, and VisualNews.  }
    \label{fig:exp1_recall10}
    \vspace{14pt} 
    \includegraphics[width=0.9\textwidth]{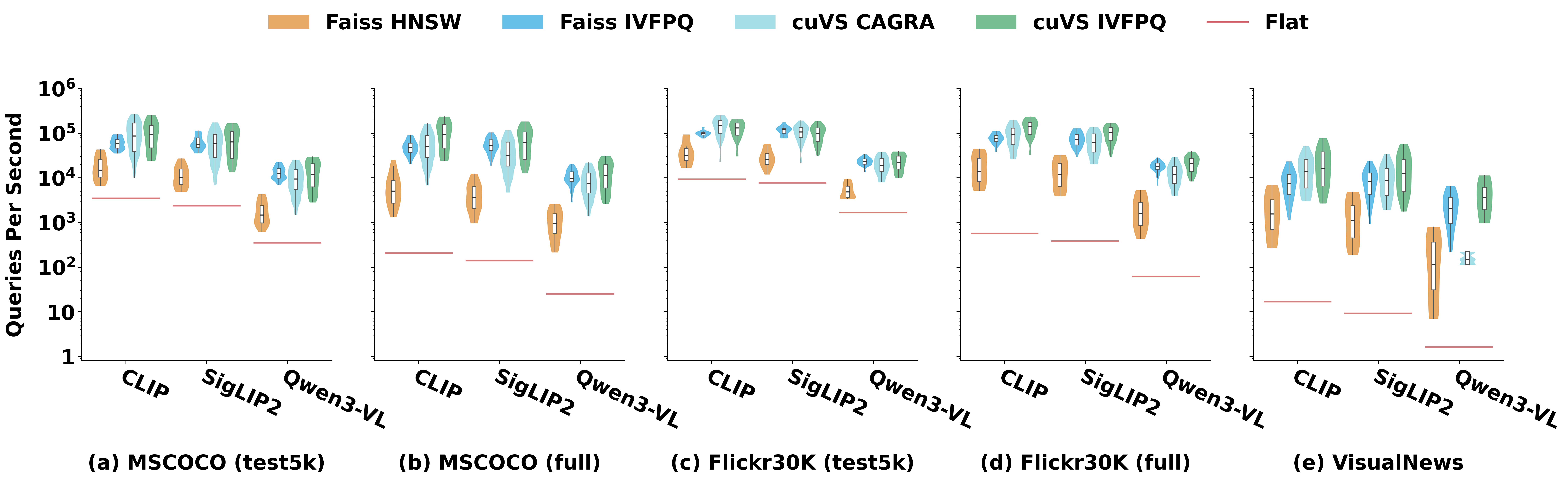}
    \caption{QPS distributions on MSCOCO, Flickr30K, and VisualNews.}
    \label{fig:exp1_qps}
\end{figure}
% zadd 狂写

We next quantify how much of the embedding's exact-search recall is retained by the ANN backend. Under matched ECs, we define $\operatorname{Gap}(e,a,\beta,\gamma,d,k) = U(e,d,k) - R_k(e,a,\beta,\gamma,d)$ and $\operatorname{Retention}(e,a,\beta,\gamma,d,k) = R_k(e,a,\beta,\gamma,d) / U(e,d,k)$ for $U(e,d,k)>0$. A Retention score near 1 indicates that ANN recall closely matches the corresponding Flat reference.

Figure~\ref{fig:exp1_recall10} shows that graph-based indexes are highly effective at retaining exact-search recall: cuVS CAGRA and Faiss HNSW achieve mean Retention@10 values of 0.9959 and 0.9875, respectively, across all configurations. In contrast, PQ-based indexes trade recall for memory efficiency, with Faiss IVFPQ dropping to 0.7932. Detailed retention and Gap-to-Flat statistics are provided in Appendix~\ref{sec:detailed_results}.

\begin{insightline}
\textbf{Insight 3. \emph{ANN indexes may lose relevant results through approximation but cannot improve the underlying embedding's semantic alignment.}}
\end{insightline}

\subsubsection{Recall-QPS frontier}

% zadd2 合并的图表>W<
% \begin{figure}[t]
%     \centering
%     \includegraphics[width=0.9\textwidth]{figure/qps_comparison_plot_violin.png}
%     \caption{QPS distributions on MSCOCO, Flickr30K, and VisualNews.}
%     \label{fig:exp1_qps}
% \end{figure}

VecDBLens further evaluates how embeddings and indexes jointly shape the Pareto frontier of retrieval quality and system cost. While Qwen3-VL provides the highest exact-search recall, searching its 4096-dimensional vectors incurs substantially lower throughput. Under Flat search, Qwen3-VL is $5.6\times$--$10.0\times$ slower than CLIP across dataset conditions.

However, ANN indexing substantially reduces the search cost while retaining retrieval quality. As shown in Figure~\ref{fig:exp1_qps}, among configurations with $\mathrm{Retention}@10 \geq 0.99$, Faiss HNSW boosts QPS from 13.8 (Flat) to 1898.6 on VisualNews with CLIP. GPU-accelerated CAGRA reaches 29,433.6 QPS while retaining 99.28\% of the corresponding Flat recall.

Ultimately, end-to-end performance depends on interactions among retrieval functions. The embedding function determines exact-search recall and influences distance-computation cost, while the indexing and query-processing functions determine the recall-throughput trade-off relative to this reference.

\begin{insightline}
  \textbf{Insight 4. \emph{The Recall-QPS Pareto frontier is jointly determined: embeddings establish the exact-search reference, and indexing and query processing determine how efficiently its recall is retained.}}
\end{insightline}

\subsubsection{Cross-object interactions}
\label{subsec:exp_module_interactions}

We examine how the performance effect of one evaluated object depends on other objects using functional ANOVA (Section~\ref{sec:distributional_metrics}).
Table~\ref{tab:interaction_summary} shows that interactions explain a larger share of Retention@10 variation than of throughput variation.
Embedding--query processing dominates IVFPQ's quality interactions.
For HNSW, the dominant contribution varies between embedding--index construction and embedding--query processing across build repetitions.
Three-factor interactions remain small for both indexes.

Conditional comparisons illustrate how query processing moderates construction effects. Increasing HNSW's \texttt{efSearch} by $8\times$ reduces the mean standard deviation of Retention@10 across construction configurations by 36.0\%--78.0\%, showing that larger search budgets can reduce sensitivity to index construction.

\begin{table*}[t]
\centering
\caption{Interaction shares of within-condition variance (\%) across five evaluation conditions. Ranges cover three build repetitions, with each run analyzed separately. $E$, $I$, and $S$ denote embedding choice, index construction, and query processing, respectively.}
\label{tab:interaction_summary}
\small
\begin{tabular}{lcccc}
\toprule
& \multicolumn{2}{c}{\textbf{Retention@10}}
& \multicolumn{2}{c}{$\boldsymbol{\log_2(\mathrm{QPS})}$} \\
\cmidrule(lr){2-3}\cmidrule(lr){4-5}
\textbf{Interaction} & \textbf{Faiss HNSW} & \textbf{Faiss IVFPQ}
            & \textbf{Faiss HNSW} & \textbf{Faiss IVFPQ} \\
\midrule
$E \times I$
&\cellcolor{vdbTeal!15} 1.19--17.89 &\cellcolor{vdbTeal!15} 0.74--2.71
&\cellcolor{vdbTeal!15} 0.26--0.33 &\cellcolor{vdbTeal!15} 0.34--1.58 \\
$E \times S$
&\cellcolor{vdbTeal!15} 3.66--16.89 &\cellcolor{vdbTeal!15} 6.72--6.99
&\cellcolor{vdbTeal!15} 0.23--0.97 &\cellcolor{vdbTeal!15} 0.82--1.16 \\
$I \times S$
&\cellcolor{vdbTeal!15} 0.30--1.69 &\cellcolor{vdbTeal!15} 0.17--0.23
&\cellcolor{vdbTeal!15} 0.03--0.20 &\cellcolor{vdbTeal!15} 0.56--0.80 \\
$E \times I \times S$
&\cellcolor{vdbTeal!15} 0.16--1.02 &\cellcolor{vdbTeal!15} 0.14--0.20
&\cellcolor{vdbTeal!15} 0.07--0.26 &\cellcolor{vdbTeal!15} 0.10--0.43 \\
\midrule
Total interactions
&\cellcolor{vdbTeal!45} 19.34--22.01 &\cellcolor{vdbTeal!45} 7.89--10.03
&\cellcolor{vdbTeal!45} 0.67--1.75 &\cellcolor{vdbTeal!45} 1.85--3.96 \\
\bottomrule
\end{tabular}
\end{table*}

\begin{insightline}
\textbf{Insight 5. \emph{Cross-object interactions explain a larger share of variation in retrieval quality than in throughput.}}
\end{insightline}

\subsection{Experiment 2: Parameterization Attribution in Filtered ANN}
\label{subsec:exp_filtered_ann}

This experiment investigates filtered ANN from three attribution perspectives. For algorithm-level comparisons, the EO is the method $a$, while parameters, filter ratio, dimensionality, and runtime settings are ECs. For parameter monotonicity analysis, the EO is a controlled parameter change (e.g., $M$ or \texttt{efConstruction}), with the remaining factors fixed as matched ECs. Parameters linked by a configuration rule are changed jointly. For workload analysis, filter ratio and dimensionality define EC strata for comparing exact pre-filtering with approximate search.

\subsubsection{Non-monotonicity of stronger parameterization}

We compare adjacent parameter levels under matched ECs and record a violation whenever Recall@100 decreases. Table~\ref{tab:monotonicity_audit} shows that both ACORN and DiGRA exhibit such decreases as construction parameters increase.

We identify two mechanisms that can reduce filtered recall in the evaluated ACORN implementation. First, increasing graph connectivity raises the threshold for exploring neighbors of neighbors. As Table~\ref{tab:acorn_connectivity} shows, nodes in the denser graph have more filter-compatible neighbors on average, but fewer nodes qualify for two-hop expansion and Recall@100 falls.

Second, a larger construction budget can allow the candidate set to reach a stopping threshold before closer candidates replace those encountered earlier. The resulting graph has longer, less directionally diverse edges. Query traces show that search on this graph visits more nodes but reaches fewer exact neighbors.

We also observe recall decreases under stronger ACORN construction settings on App Reviews and in both CC-News time windows. Appendix~\ref{app:filtered_parameterization} reports the real-workload results.

% We first audit whether stronger parameterization strictly improves recall. Fixing all other ECs, we compare adjacent parameter levels and record a violation if increasing the budget yields a lower Recall@100.
% Table~\ref{tab:monotonicity_audit} indicates that increasing query-time search budgets generally enters a saturated regime with negligible regressions. However, the performance of universal graph parameters $M$ and $efC$ is highly conditional on filter selectivity. For ACORN, increasing \texttt{efConstruction} triggers 306 violations, with recall drops up to 0.5237. Our analysis shows these $efC$ regressions are most pervasive under loose filtering ($r=50\%$), where the violation rate reaches 98.8\%.

% Expanding graph connectivity ($M$) also frequently disrupts the filtered topology, causing 273 violations with a maximum drop of 0.4332. Notably, $M$ exhibits its highest instability under highly selective filters ($r \le 0.01\%$), where increasing connectivity often introduces "noisy" edges that dilute valid search paths.

% These results reveal that filtered ANN parameterization is conditional: while search budgets yield diminishing returns, aggressive offline connectivity ($M$) and construction budget ($efC$) can arbitrarily alter the effective neighborhood structure under active filters, leading to severe recall regression.

\begin{table}[t]
\centering
\caption{Matched-condition monotonicity audit for filtered ANN parameterization. Each comparison increases the selected parameter while keeping other settings fixed, except for parameters coupled by a configuration rule. A violation is recorded when Recall@100 decreases.}
\label{tab:monotonicity_audit}
\small
\begin{tabular}{lrrrr}
\toprule
\textbf{Parameter change} & \textbf{Comparisons} & \textbf{Violations} & \textbf{Rate} & \textbf{Max drop} \\
\midrule
ACORN: $M$ & 2244 & 273 &\cellcolor{vdbCoral!18} 12.2\% &\cellcolor{vdbCoral!45} 0.4332 \\
ACORN: \texttt{efConstruction} & 1683 & 306 &\cellcolor{vdbCoral!25} 18.2\% &\cellcolor{vdbCoral!53} 0.5237 \\
ACORN: $\gamma_{\mathrm{ACORN}}$ & 1836 & 762 &\cellcolor{vdbCoral!51} 41.5\% &\cellcolor{vdbCoral!11} 0.0626 \\
DiGRA: $M$ & 520 & 133 &\cellcolor{vdbCoral!33} 25.6\% &\cellcolor{vdbCoral!6} 0.010891 \\
\bottomrule
\end{tabular}
% \begin{tabular}{lccccc}
% \toprule
% \textbf{Parameter}
% & \textbf{Pairs}
% & \textbf{Violations}
% & \textbf{Violation rate}
% & \textbf{Max drop}
% & \textbf{Mean drop} \\
% \midrule
% ACORN\_$efS$
% & 3168 & 0 & 0.0\%
% & 0 & 0 \\
% DiGRA\_$ef$
% & 480 & 33 & 6.9\%
% & $1.90{\times}10^{-5}$ & $<$ $2{\times}10^{-5}$ \\
% RangePQ\_$l$
% & 1440 & 0 & 0.0\%
% & 0 & 0 \\
% \midrule
% ACORN\_$m$
% & 2244
% & \textcolor{blue}{{273}}
% & \textcolor{blue}{12.2\%}
% & {0.4332}
% & {0.1112} \\
% ACORN\_$efC$
% & 1683
% & {306}
% & {18.2\%}
% & \textcolor{red}{\underline{0.5237}}
% & \textcolor{red}{\underline{0.1518}} \\
% ACORN\_$gamma$
% & 1836
% & \textcolor{red}{\underline{762}}
% & \textcolor{red}{\underline{41.5\%}}
% & \textcolor{blue}{0.0626}
% & \textcolor{blue}{0.0085} \\
% \bottomrule
% \end{tabular}
\end{table}

\begin{table}[t]
\centering
\caption{Effect of increased graph connectivity on ACORN under matched evaluation conditions.}
\label{tab:acorn_connectivity}
\small
\begin{tabular}{lrr}
\toprule
\textbf{Measure} & \textbf{Baseline graph} & \textbf{Denser graph} \\
\midrule
Mean number of filter-compatible neighbors &\cellcolor{vdbTeal!12} 0.1596 &\cellcolor{vdbTeal!35} 0.2002 \\
Nodes eligible for two-hop expansion (\%) &\cellcolor{vdbTeal!12} 59.97 &\cellcolor{vdbCoral!35} 18.78 \\
Recall@100 &\cellcolor{vdbTeal!12} 0.3150 &\cellcolor{vdbCoral!35} 0.1827 \\
\bottomrule
\end{tabular}
\end{table}

% \begin{insightbox}
% \finding{Insight 5. Stronger index construction parameterization is not monotonically beneficial; aggressive graph connectivity can disrupt filtered neighborhoods and regress recall.}
% \end{insightbox}

\begin{insightline}
\textbf{Insight 6. \emph{Stronger index construction parameterization is not monotonically beneficial; aggressive graph connectivity can disrupt filtered neighborhoods and regress recall.}}
\end{insightline}

% zadd2 recall-qps图

\subsubsection{Effect of filter ratio and dimensionality}

\begin{figure}[htbp]
    \centering
    \includegraphics[width=1\textwidth]{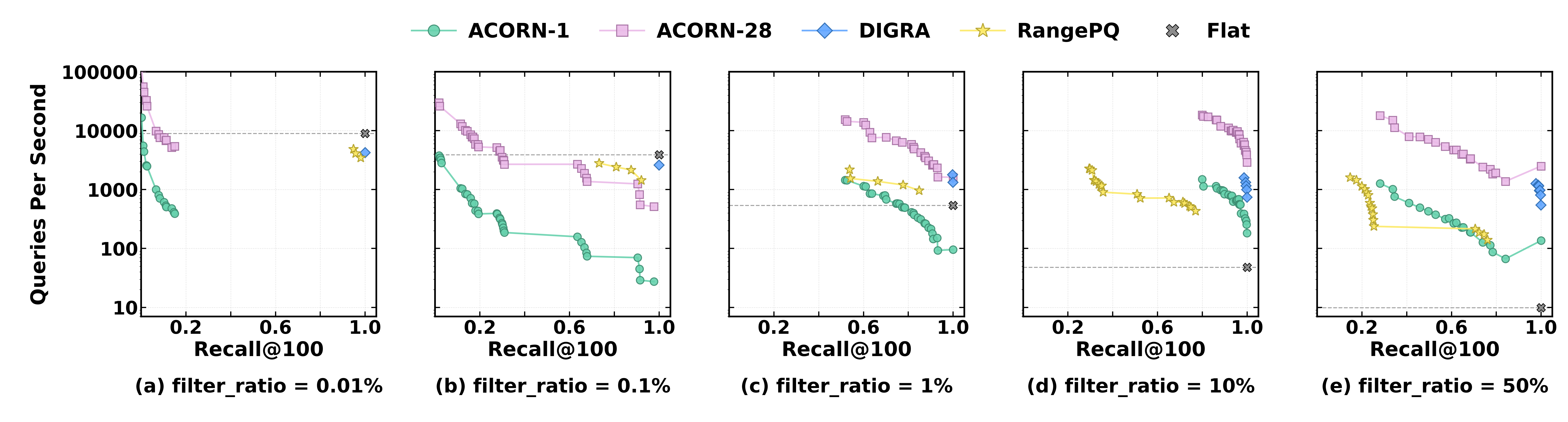}
    \caption{Recall-QPS Pareto frontiers across various filter ratios (128-dimensional data).}
    \label{fig:recall_qps_vec128}
    
    \vspace{10pt}
    \includegraphics[width=1\textwidth]{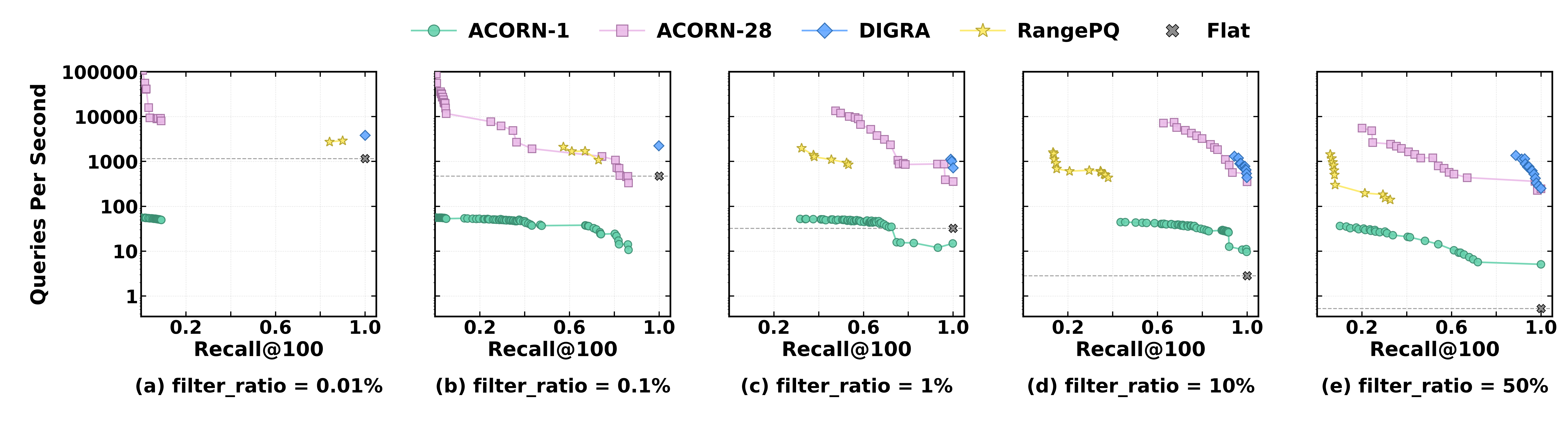}
    \caption{Recall-QPS Pareto frontiers across various filter ratios (960-dimensional data).}
    \label{fig:recall_qps_vec960}

\end{figure}

We examine how filter ratio and dimensionality affect the choice between exact pre-filtering and approximate search. On the 128-dimensional SIFT workload (Figure~\ref{fig:recall_qps_vec128}), Flat exact pre-filtering remains competitive at filter ratios of 0.01\% and 0.1\% because few vectors satisfy the filter. At ratios of 1\% and above, Flat must scan a larger candidate set, and DiGRA achieves the highest single-threaded QPS at a target Recall@100 of $\tau=0.95$.

On the 960-dimensional GIST workload (Figure~\ref{fig:recall_qps_vec960}), each distance computation is more expensive. DiGRA achieves the highest single-threaded QPS at the same recall target across all tested filter ratios, including 0.01\%. For filter ratios from 1\% to 50\%, ACORN-28 achieves $10.8\times$--$1051.7\times$ the QPS of single-threaded Flat on SIFT and $97.4\times$--$1842.9\times$ on GIST.

\begin{insightline}
\textbf{Insight 7. \emph{The boundary between exact pre-filtering and approximate search is non-static; it shifts substantially based on the joint effect of vector dimensionality and filter selectivity.}}   
\end{insightline}

% \begin{insightbox}
% \finding{Insight 6. The boundary between exact pre-filtering and approximate search is non-static; it shifts substantially based on the joint effect of vector dimensionality and filter selectivity.}
% \end{insightbox}

\subsubsection{Runtime and cost regimes}

\begin{figure}[htbp]
    \centering
    \includegraphics[width=0.8\linewidth]{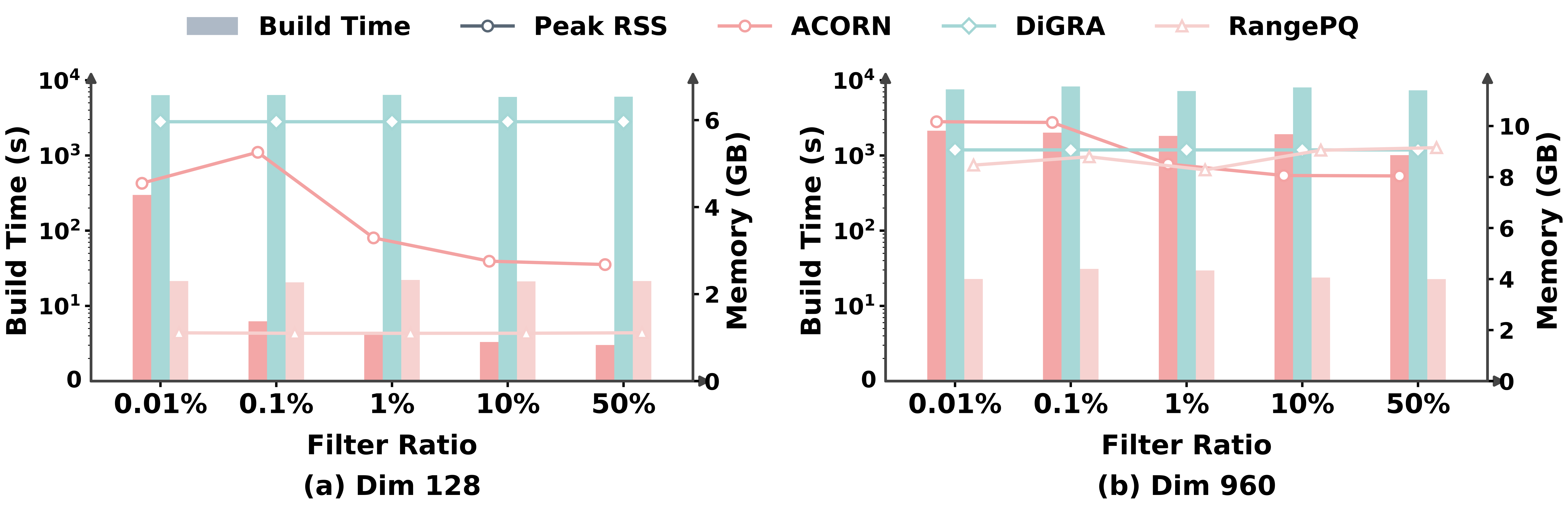}
    \caption{
 Construction time and memory use of filtered ANN methods across filter ratios and dimensionalities.
    }
    \label{fig:buildtime_peakrss}
\end{figure}

Beyond recall and throughput, practical deployment depends on index construction time and memory use. Figure~\ref{fig:buildtime_peakrss} shows how these costs vary across filtered ANN methods. DiGRA achieves near-perfect recall in many configurations but requires long build times and substantial memory. RangePQ builds quickly and has the smallest increase in memory use during construction, but none of its tested configurations reaches the Recall@100 target of 0.95 on 960-dimensional GIST. Meanwhile, ACORN strikes a balance, achieving robust Recall-QPS trade-offs—especially when multi-threading (ACORN-28) yields a $14.2\times$ median QPS speedup.
Consequently, filtered ANN performance cannot be abstracted to a single metric. It requires object-level diagnosis across recall, throughput, build time, and parameter stability.

\begin{insightline}
\textbf{Insight 8. \emph{Distinct algorithmic cost regimes reshape the attainable Pareto frontier; optimal algorithm selection must balance memory, build time, and multi-threading efficiency.} }  
\end{insightline}

\section{Conclusion}
This paper introduces \textbf{VecDBLens}, an Evaluatology-based diagnostic framework that shifts vector database retrieval pipeline evaluation from \textbf{black-box database comparison} to \textbf{object attribution} across \textbf{\emph{embedding, indexing, and query processing}} functions. Our analyses yield two critical insights. First, the embedding space strictly defines the upper bound of retrieval performance; downstream ANN indexes only dictate the efficiency-cost trade-off in realizing this potential, rather than creating semantic relevance. Second, in filtered search, stronger index parameterization is not monotonically beneficial and can surprisingly regress recall. Furthermore, the optimal boundary between exact pre-filtering and ANN approximation is non-static, jointly governed by dimensionality and filter selectivity. Ultimately, by isolating bottlenecks to representation limits, index approximations, or query executions, VecDBLens prevents misdirected system optimizations and guides more robust vector database design. 

\section{Limitations and Future Work}
\label{app:limitations}

VecDBLens currently emphasizes recall, throughput, memory usage, construction cost, and parameter stability. These metrics characterize retrieval quality and system cost but provide limited insight into semantic failure mechanisms. Future work will develop finer-grained semantic diagnostics and incorporate energy efficiency and incremental indexing costs.

Beyond cross-modal and filtered retrieval, we have extended evaluation to Chinese text retrieval and multi-vector retrieval in scientific and financial domains (Appendix~\ref{app:additional_scenarios}). Broader language coverage, temporal data streams, and dynamic workloads remain directions for further evaluation. Extending attribution to database functions such as update processing and query optimization would also broaden the framework's scope.

We are committed to long-term maintenance, including continued updates to the implementation, evaluation protocols, and workload coverage as vector retrieval evolves. Our goal is to advance evaluation from asking which system ranks first to identifying which retrieval function limits performance, under which conditions, and why. This perspective supports reproducible research, efficient system development, and application-aware vector database design.

\section*{Acknowledgments and Disclosures}

%We thank the anonymous reviewers for carefully reading our manuscript and providing constructive feedback that helped improve the clarity and presentation of this paper. This work was supported by the project ``Research on Automatic Design Methods for Key Components of Computing Power Networks Based on Evaluatology'' (Project No.~E661110000). The views and opinions expressed in this paper are those of the authors and do not necessarily reflect those of the supporting organization. The authors declare no competing interests related to this work.

We thank the anonymous reviewers for carefully reading our manuscript and providing constructive feedback that helped improve the clarity and presentation of this paper. This work was supported by the Innovation Funding of the Institute of Computing Technology, Chinese Academy of Sciences (ICT, CAS), under Grant No. E661110.

\bibliographystyle{unsrt} % Or plainnat, unsrt, etc.
\bibliography{ref} % Points to my_references.bib

@misc{qdrant,
  author = {{Qdrant Team}},
  title = {Qdrant: Open Source Vector Database and Search Engine},
  year = {2021},
  publisher = {GitHub},
  journal = {GitHub repository},
  howpublished = {\url{https://github.com/qdrant/qdrant}}
}

@article{rangepq,
author = {Zhang, Fangyuan and Jiang, Mengxu and Hou, Guanhao and Shi, Jieming and Fan, Hua and Zhou, Wenchao and Li, Feifei and Wang, Sibo},
title = {Efficient Dynamic Indexing for Range Filtered Approximate Nearest Neighbor Search},
year = {2025},
issue_date = {June 2025},
publisher = {Association for Computing Machinery},
address = {New York, NY, USA},
volume = {3},
number = {3},
url = {https://doi.org/10.1145/3725401},
doi = {10.1145/3725401},
journal = {Proc. ACM Manag. Data},
month = jun,
articleno = {152},
numpages = {26}
}

@inproceedings{wang2021milvus,
  title={Milvus: A purpose-built vector data management system},
  author={Wang, Jianguo and Yi, Xiaomeng and Guo, Rentong and Jin, Hai and Xu, Peng and Li, Shengjun and Wang, Xiangyu and Guo, Xiangzhou and Li, Chengming and Xu, Xiaohai and others},
  booktitle={Proceedings of the 2021 international conference on management of data},
  pages={2614--2627},
  year={2021}
}

@article{johnson2019billion,
  title={Billion-scale similarity search with {GPU}s},
  author={Johnson, Jeff and Douze, Matthijs and J{\'e}gou, Herv{\'e}},
  journal={IEEE Transactions on Big Data},
  volume={7},
  number={3},
  pages={535--547},
  year={2021},
  publisher={IEEE}
}

@article{malkov2018efficient,
  title={Efficient and robust approximate nearest neighbor search using hierarchical navigable small world graphs},
  author={Malkov, Yu A. and Yashunin, D. A.},
  journal={IEEE Transactions on Pattern Analysis and Machine Intelligence},
  volume={42},
  number={4},
  pages={824--836},
  year={2020},
  publisher={IEEE}
}

@article{jegou2011product,
  title={Product quantization for nearest neighbor search},
  author={J{\'e}gou, Herv{\'e} and Douze, Matthijs and Schmid, Cordelia},
  journal={IEEE Transactions on Pattern Analysis and Machine Intelligence},
  volume={33},
  number={1},
  pages={117--128},
  year={2011},
  publisher={IEEE}
}

@inproceedings{radford2021clip,
  title={Learning transferable visual models from natural language supervision},
  author={Radford, Alec and Kim, Jong Wook and Hallacy, Chris and Ramesh, Aditya and Goh, Gabriel and Agarwal, Sandhini and Sastry, Girish and Askell, Amanda and Mishkin, Pamela and Clark, Jack and others},
  booktitle={Proceedings of the 38th International Conference on Machine Learning},
  pages={8748--8763},
  year={2021}
}

@article{tschannen2025siglip2,
  title={{SigLIP} 2: Multilingual vision-language encoders with improved semantic understanding, localization, and dense features},
  author={Tschannen, Michael and Gritsenko, Alexey and Wang, Xiao and Naeem, Muhammad Ferjad and Alabdulmohsin, Ibrahim and Parthasarathy, Nikhil and Evans, Talfan and Beyer, Lucas and Xia, Ye and Mustafa, Basil and others},
  journal={arXiv preprint arXiv:2502.14786},
  year={2025}
}

@article{bai2025qwen3vl,
  title={Qwen3-{VL} Technical Report},
  author={Bai, Shuai and others},
  journal={arXiv preprint arXiv:2511.21631},
  year={2025}
}

@misc{hnswlib,
  title={{hnswlib} -- Header-only {C++}/python library for fast approximate nearest neighbors},
  author={Malkov, Yury and others},
  howpublished={\url{https://github.com/nmslib/hnswlib}},
  year={2018}
}

@article{patel2024acorn,
  title={{ACORN}: Performant and predicate-agnostic search over vector embeddings and structured data},
  author={Patel, Liana and Kraft, Peter and Guestrin, Carlos and Zaharia, Matei},
  journal = {Proc. ACM Manag. Data},
  volume = {2},
  number = {3},
  articleno = {120},
  year = {2024},
  doi = {10.1145/3654923}
}

@article{jiang2025digra,
  title={{DIGRA}: A dynamic graph indexing for approximate nearest neighbor search with range filter},
  author={Jiang, Mengxu and Yang, Zhi and Zhang, Fangyuan and Hou, Guanhao and Shi, Jieming and Zhou, Wenchao and Li, Feifei and Wang, Sibo},
  journal={Proc. ACM Manag. Data},
  year={2025}
}

@article{young2014flickr30k,
  title={From image descriptions to visual denotations: New similarity metrics for semantic inference over event descriptions},
  author={Young, Peter and Lai, Alice and Hodosh, Micah and Hockenmaier, Julia},
  journal={Transactions of the Association for Computational Linguistics},
  volume={2},
  pages={67--78},
  year={2014}
}

@inproceedings{lin2014mscoco,
  title={{Microsoft COCO}: Common objects in context},
  author={Lin, Tsung-Yi and Maire, Michael and Belongie, Serge and Hays, James and Perona, Pietro and Ramanan, Deva and Doll{\'a}r, Piotr and Zitnick, C. Lawrence},
  booktitle={Computer Vision -- ECCV 2014},
  pages={740--755},
  year={2014},
  publisher={Springer International Publishing}
}

@inproceedings{liu2021visualnews,
  title={{VisualNews}: Benchmark and challenges in news image captioning},
  author={Liu, Fuxiao and Wang, Yinghan and Wang, Tianlu and Ordonez, Vicente},
  booktitle={Proceedings of the 2021 Conference on Empirical Methods in Natural Language Processing},
  year={2021}
}

@inproceedings{ootomo2023cagra,
  title={{CAGRA}: Highly parallel graph construction and approximate nearest neighbor search for {GPU}s},
  author={Ootomo, Hiroyuki and Naruse, Akira and Nolet, Corey and Wang, Ray and Feher, Tamas and Wang, Yong},
  booktitle={2024 IEEE 40th International Conference on Data Engineering (ICDE)},
  year={2024}
}

@inproceedings{jegou2010sift,
  title={Searching in one billion vectors: Re-rank with source coding},
  author={J{\'e}gou, Herv{\'e} and Tavenard, Romain and Douze, Matthijs and Amsaleg, Laurent},
  booktitle={2011 IEEE International Conference on Acoustics, Speech and Signal Processing (ICASSP)}, 
  year={2011},
  pages = {861--864},
  doi = {10.1109/ICASSP.2011.5946540}
}

@article{zhan2024evaluatology,
  title={Evaluatology: The science and engineering of evaluation},
  author={Zhan, Jianfeng and Wang, Lei and Gao, Wanling and Li, Hongxiao and Wang, Chenxi and Huang, Yunyou and Li, Yatao and Yang, Zhengxin and Kang, Guoxin and Luo, Chunjie and others},
  journal={BenchCouncil Transactions on Benchmarks, Standards and Evaluations},
  volume={4},
  number={1},
  pages={100162},
  year={2024},
  publisher={Elsevier}
}

@article{benchmark,
  title = {ANN-Benchmarks: A benchmarking tool for approximate nearest neighbor algorithms},
  journal = {Information Systems},
  volume = {87},
  pages = {101374},
  year = {2020},
  author = {Martin Aumüller and Erik Bernhardsson and Alexander Faithfull}
}

@book{zhan2025evaluatology,
  title     = {Evaluatology: The Science of Uncovering the Effects},
  author    = {Zhan, Jianfeng and Wang, Lei and Gao, Wanling and Wang, Chenxi and Li, Hongxiao and Fan, Fanda and Kang, Guoxin},
  year      = {2025},
  publisher = {BenchCouncil Press},
  address   = {Hong Kong},
  isbn      = {978-988-71596-8-1},
  edition   = {Version 1.0 (First Edition, 2025)}
}

@article{wang2025achieving,
  author = {Wang, Chenxi and Wang, Lei and Gao, Wanling and Fan, Fanda and Su, Yuchen and Zhou, Yutong and Yang, Yikang and Zhan, Jianfeng},
  title = {Achieving Consistent and Comparable {CPU} Evaluation},
  journal = {arXiv preprint arXiv:2411.08494},
  year = {2025}
}

@inproceedings{tianbeyond,
  title={Beyond Benchmarks: Toward Causally Faithful Evaluation of Large Language Models},
  author={Tian, Zhengshuyuan and Gao, Wanling and Lan, Chuanxin and Wang, Chenxi and Wang, Lei and Kang, Guoxin and Yang, Zhengxin and Huang, Yunyou and Hong, Xuehai and Zhan, Jianfeng},
  booktitle={Forty-third International Conference on Machine Learning}
}

@inproceedings{
wang2026combinationts,
title={Combination{TS}: A Modular Framework for Understanding Time-Series Forecasting Models},
author={Xiaorui Wang and Fanda Fan and Chenxi Wang and Yuxuan Yang and Rui Tang and Kuoyu Gao and simiao pang and Yuanfeng Shang and Zhipeng Liu and Wanling Gao and Lei Wang and Jianfeng Zhan},
booktitle={Forty-third International Conference on Machine Learning},
year={2026},
url={https://openreview.net/forum?id=CwHRT46VmC}
}

@article{BigVectorBench,
  title={BigVectorBench: Heterogeneous Data Embedding and Compound Queries are Essential in Evaluating Vector Databases},
  author={Kang, Guoxin and Ge, Zhongxin and Hu, Jingpei and Zhang, Xueya and Wang, Lei and Zhan, Jianfeng},
  journal={Proceedings of the VLDB Endowment},
  volume={18},
  number={5},
  pages={1536--1550},
  year={2025},
  publisher={VLDB Endowment}
}

@misc{big-ann-benchmarks,
  title = {big-ann-benchmarks: {Framework} for evaluating {ANNS} algorithms on billion scale datasets.},
  url = {https://github.com/harsha-simhadri/big-ann-benchmarks},
  author = {Harsha Vardhan Simhadri},
  year = {2023}
}

@article{10.1007/s00778-024-00864-x,
  author = {Pan, James Jie and Wang, Jianguo and Li, Guoliang},
  title = {Survey of vector database management systems},
  journal={arXiv preprint arXiv:2310.14021},
  year={2023}
}

@article{gao2023retrieval,
  title={Retrieval-Augmented Generation for Large Language Models: A Survey},
  author={Gao, Yunfan},
  journal={arXiv preprint arXiv:2312.10997},
  year={2023}
}

@ARTICLE{8681160,
  author={Li, Wen and Zhang, Ying and Sun, Yifang and Wang, Wei and Li, Mingjie and Zhang, Wenjie and Lin, Xuemin},
  journal={IEEE Transactions on Knowledge and Data Engineering}, 
  title={Approximate Nearest Neighbor Search on High Dimensional Data — Experiments, Analyses, and Improvement}, 
  volume={32},
  number={8},
  pages={1475--1488},
  year={2019},
  publisher={IEEE}
}

@misc{wang2021comprehensivesurveyexperimentalcomparison,
  title={A Comprehensive Survey and Experimental Comparison of Graph-Based Approximate Nearest Neighbor Search}, 
  author={Wang, Mengzhao and Xu, Xiaoliang and Yue, Qiang and Wang, Yuxiang},
  journal={arXiv preprint arXiv:2101.12631},
  year={2021}
}

@article{10.14778/3554821.3554843,
  title={Manu: a cloud native vector database management system},
  author={Guo, Rentong and Luan, Xiaofan and Xiang, Long and Yan, Xiao and Yi, Xiaomeng and Luo, Jigao and Cheng, Qianya and Xu, Weizhi and Luo, Jiarui and Liu, Frank and others},
  journal={arXiv preprint arXiv:2206.13843},
  year={2022}
}

@article{10.14778/3685800.3685805,
  title={Singlestore-v: An integrated vector database system in singlestore},
  author={Chen, Cheng and Jin, Chenzhe and Zhang, Yunan and Podolsky, Sasha and Wu, Chun and Wang, Szu-Po and Hanson, Eric and Sun, Zhou and Walzer, Robert and Wang, Jianguo},
  journal={Proceedings of the VLDB Endowment},
  volume={17},
  number={12},
  year={2024},
  publisher={VLDB Endowment}
}

@inproceedings{10.1145/276698.276876,
  title={Approximate nearest neighbors: towards removing the curse of dimensionality},
  author={Indyk, Piotr and Motwani, Rajeev},
  booktitle={Proceedings of the thirtieth annual ACM symposium on Theory of computing},
  pages={604--613},
  year={1998}
}

@article{10.1145/3639324,
author = {Zuo, Chaoji and Qiao, Miao and Zhou, Wenchao and Li, Feifei and Deng, Dong},
title = {SeRF: Segment Graph for Range-Filtering Approximate Nearest Neighbor Search},
  journal={Proceedings of the ACM on Management of Data},
  volume={2},
  number={1},
  pages={1--26},
  year={2024},
  publisher={ACM New York, NY, USA}
}

@inproceedings{10.5555/3692070.3692567,
  author = {Engels, Joshua and Landrum, Benjamin and Yu, Shangdi and Dhulipala, Laxman and Shun, Julian},
  title = {Approximate nearest neighbor search with window filters},
  journal={arXiv preprint arXiv:2402.00943},
  year={2024}
}

@article{10.1145/3698814,
  author = {Xu, Yuexuan and Gao, Jianyang and Gou, Yutong and Long, Cheng and Jensen, Christian S.},
  title = {iRangeGraph: Improvising Range-dedicated Graphs for Range-filtering Nearest Neighbor Search},
  journal={Proceedings of the ACM on Management of Data},
  volume={2},
  number={6},
  pages={1--26},
  year={2024},
  publisher={ACM New York, NY, USA}
}

@article{10.14778/3717755.3717770,
  author = {Liang, Anqi and Zhang, Pengcheng and Yao, Bin and Chen, Zhongpu and Song, Yitong and Cheng, Guangxu},
  title = {UNIFY: Unified Index for Range Filtered Approximate Nearest Neighbors Search},
  journal={arXiv preprint arXiv:2412.02448},
  year={2024}
}

@article{10.1145/3769843,
  author = {Wang, Ziqi and Zhang, Jingzhe and Hu, Wei},
  title = {WoW: A Window-to-Window Incremental Index for Range-Filtering Approximate Nearest Neighbor Search},
  journal={Proceedings of the ACM on Management of Data},
  volume={3},
  number={6},
  pages={1--27},
  year={2025},
  publisher={ACM New York, NY, USA}
}

@inproceedings{10.1145/3543507.3583552,
  title={Filtered-diskann: Graph algorithms for approximate nearest neighbor search with filters},
  author={Gollapudi, Siddharth and Karia, Neel and Sivashankar, Varun and Krishnaswamy, Ravishankar and Begwani, Nikit and Raz, Swapnil and Lin, Yiyong and Zhang, Yin and Mahapatro, Neelam and Srinivasan, Premkumar and others},
  booktitle={Proceedings of the ACM Web Conference 2023},
  pages={3406--3416},
  year={2023}
}

@article{oliva2001modeling,
  title={Modeling the shape of the scene: A holistic representation of the spatial envelope},
  author={Oliva, Aude and Torralba, Antonio},
  journal={International journal of computer vision},
  volume={42},
  number={3},
  pages={145--175},
  year={2001},
  publisher={Springer}
}

@article{10.1145/3786690,
author = {Chen, Tingyang and Fu, Cong and Wu, Jiahua and Wu, Haotian and Fan, Hua and Ke, Xiangyu and Gao, Yunjun and Ni, Yabo and Zeng, Anxiang},
title = {Reveal Hidden Pitfalls and Navigate Next Generation of Vector Similarity Search from Task-Centric Views: [Experiments \& Analysis]},
year = {2026},
issue_date = {February 2026},
publisher = {Association for Computing Machinery},
address = {New York, NY, USA},
volume = {4},
number = {1},
url = {https://doi.org/10.1145/3786690},
doi = {10.1145/3786690},
journal = {Proc. ACM Manag. Data},
month = apr,
articleno = {76},
numpages = {25}
}

@article{sobol1990sensitivity,
  title={On sensitivity estimation for nonlinear mathematical models},
  author={Sobol', Il'ya Meerovich},
  journal={Matematicheskoe modelirovanie},
  volume={2},
  number={1},
  pages={112--118},
  year={1990},
  publisher={Russian Academy of Sciences, Branch of Mathematical Sciences}
}

@book{enderton1977elements,
  title={Elements of set theory},
  author={Enderton, Herbert B},
  year={1977},
  publisher={Gulf Professional Publishing}
}

@article{wang2026poster,
  title={POSTER: BenchCPU: Evaluatology-Based CPU Benchmarking},
  author={Wang, Chenxi and Su, Yuchen and Wang, Lei and Kang, Guoxin and Gao, Wanling and Zhang, Fan and Zhan, Jianfeng},
  year={2026}
}

@article{zhang2023miracl,
  title={Miracl: A multilingual retrieval dataset covering 18 diverse languages},
  author={Zhang, Xinyu and Thakur, Nandan and Ogundepo, Odunayo and Kamalloo, Ehsan and Alfonso-Hermelo, David and Li, Xiaoguang and Liu, Qun and Rezagholizadeh, Mehdi and Lin, Jimmy},
  journal={Transactions of the Association for Computational Linguistics},
  volume={11},
  pages={1114--1131},
  year={2023},
  publisher={MIT Press, One Broadway, 12th Floor, Cambridge, Massachusetts 02142, USA~…}
}

@inproceedings{santhanam2022colbertv2,
  title={Colbertv2: Effective and efficient retrieval via lightweight late interaction},
  author={Santhanam, Keshav and Khattab, Omar and Saad-Falcon, Jon and Potts, Christopher and Zaharia, Matei},
  booktitle={Proceedings of the 2022 Conference of the North American Chapter of the Association for Computational Linguistics: Human Language Technologies},
  pages={3715--3734},
  year={2022}
}

@article{thakur2021beir,
  title={Beir: A heterogenous benchmark for zero-shot evaluation of information retrieval models},
  author={Thakur, Nandan and Reimers, Nils and R{\"u}ckl{\'e}, Andreas and Srivastava, Abhishek and Gurevych, Iryna},
  journal={arXiv preprint arXiv:2104.08663},
  year={2021}
}

@inproceedings{santhanam2022plaid,
  title={PLAID: an efficient engine for late interaction retrieval},
  author={Santhanam, Keshav and Khattab, Omar and Potts, Christopher and Zaharia, Matei},
  booktitle={Proceedings of the 31st ACM International Conference on Information \& Knowledge Management},
  pages={1747--1756},
  year={2022}
}

@inproceedings{khattab2020colbert,
  title={Colbert: Efficient and effective passage search via contextualized late interaction over bert},
  author={Khattab, Omar and Zaharia, Matei},
  booktitle={Proceedings of the 43rd International ACM SIGIR conference on research and development in Information Retrieval},
  pages={39--48},
  year={2020}
}

@inproceedings{wadden2020fact,
  title={Fact or fiction: Verifying scientific claims},
  author={Wadden, David and Lin, Shanchuan and Lo, Kyle and Wang, Lucy Lu and van Zuylen, Madeleine and Cohan, Arman and Hajishirzi, Hannaneh},
  booktitle={Proceedings of the 2020 conference on empirical methods in natural language processing (EMNLP)},
  pages={7534--7550},
  year={2020}
}

@article{maia201818,
  title={Www'18 open challenge: financial opinion mining and question answering},
  author={Maia, Macedo and Handschuh, Siegfried and Freitas, Andr{\'e} and Davis, Brian and McDermott, Ross and Zarrouk, Manel and Balahur, Alexandra},
  year={2018},
  publisher={Association for Computing Machinery}
}

\clearpage
\appendix

\section{Notation and Definitions}
\label{appendix:notation}

This appendix defines the evaluation condition space, individual-object attribution metrics, and cross-object interaction analysis used in Section~\ref{sec:distributional_metrics}. Table~\ref{tab:notation} summarizes the notation.

\begin{table}[htbp]
\centering
\small
\caption{Main notation used in VecDBLens.}
\label{tab:notation}
\begin{tabular}{@{}>{\columncolor{white}[0pt][\tabcolsep]}l>{\columncolor{white}[\tabcolsep][0pt]}p{0.67\linewidth}@{}}
\toprule
\textbf{Notation} & \textbf{Definition} \\
\midrule
\midrule
\rowcolor{vdbGreen!35}$d$ & Dataset instance specifying the retrieval corpus and query set. \\
$\operatorname{corpus}(d),\mathcal{Q}_d,q$ & Corpus and query set for $d$, and a query $q\in\mathcal{Q}_d$. \\
\rowcolor{vdbGreen!35}dim & Vector dimensionality. \\
$e,a$ & Embedding model and index algorithm. \\
\rowcolor{vdbGreen!35}$\beta,\gamma$ & Index construction and query-processing configurations. \\
$E_e,I_{a,\beta},S_\gamma$ & Embedding, index construction, and query-processing functions. \\
\rowcolor{vdbGreen!35}$k$ & Retrieval cutoff. \\
$\mathrm{eo},\mathrm{ec}$ & Evaluated object instance and evaluation condition instance. \\
\rowcolor{vdbGreen!35}$\Omega_{\neg\mathrm{eo}}$ & Compatible condition space for EO $\mathrm{eo}$. \\
$\mathcal{S},\mathcal{C}_s^{\mathrm{eo}}$ & Strata and conditions evaluated for $\mathrm{eo}$ in stratum $s$. \\
\rowcolor{vdbGreen!35}$M(\mathrm{eo},\mathrm{ec})$ & Performance metric measured under evaluation condition instance $\mathrm{ec}$. \\
$\operatorname{mean}_M,\operatorname{Var}_M$ & Stratified mean and average within-stratum variance. \\
\rowcolor{vdbGreen!35}$\hat\Delta_{u-v}$ & Mean performance difference between EOs $u$ and $v$ under shared conditions. \\
$R_k(\mathrm{eo},\mathrm{ec}),U(e,d,k)$ & Evaluated recall and the corresponding Flat recall. \\
\rowcolor{vdbGreen!35}$E,I,S$ & Embedding, construction, and query-processing factors in the interaction analysis. \\
$y_{\mathrm{ec}}, f_{u,\mathrm{ec}}$ & Grid response and effect associated with factor subset $u$. \\
\rowcolor{vdbGreen!35}$V_{u,\mathrm{ec}},\rho_u$ & Effect variance within evaluation condition instance $\mathrm{ec}$ and its aggregated variance share. \\
\bottomrule
\end{tabular}
\end{table}

\paragraph{Evaluation condition space.}
Let $\mathcal{G}$ contain the evaluation factors: dataset instance (specifying the corpus and query set), embedding, index algorithm, construction configuration, query-processing configuration, retrieval cutoff, workload, and runtime environment. Each factor $g$ has a set of admissible values $\Omega_g$. For an EO $\mathrm{eo}$, let $\mathcal{G}_{\mathrm{eo}}$ denote the factors specified by that object. Its condition space is
\begin{equation}
\Omega_{\neg\mathrm{eo}}
=\left\{\mathrm{ec}\in\prod_{g\in\mathcal{G}\setminus\mathcal{G}_{\mathrm{eo}}}\Omega_g:
(\mathrm{eo},\mathrm{ec})\text{ forms a compatible retrieval configuration}\right\}.
\label{eq:app_condition_space}
\end{equation}
An embedding EO specifies the embedding factor; an index-configuration EO may specify both the algorithm and its construction settings. Index-specific query-processing mechanisms are evaluated within their corresponding index families. In Experiment~1, the dataset instance $d$ is one part of the evaluation condition. Each observation $M(\mathrm{eo},\mathrm{ec})$ is measured from the resulting complete configuration. Algorithm-level efficiency comparisons share the framework, hardware, and thread settings.

\subsection{Formal Definitions of Attribution Metrics and Interactions}
\label{sec:metric_definitions}

\subsubsection{Individual Object Attribution}

\paragraph{Mean performance and stability.}
Partition the evaluated conditions into strata $\mathcal{S}$, such as datasets or filter-ratio groups. For nonempty condition sets $\mathcal{C}_s^{\mathrm{eo}}$, define
\begin{align}
\operatorname{mean}_M(\mathrm{eo})
&=\frac{1}{|\mathcal{S}|}\sum_{s\in\mathcal{S}}
\frac{1}{|\mathcal{C}_s^{\mathrm{eo}}|}\sum_{\mathrm{ec}\in\mathcal{C}_s^{\mathrm{eo}}}M(\mathrm{eo},\mathrm{ec}),
\label{eq:app_stratified_mean}\\
\operatorname{Var}_M(\mathrm{eo})
&=\frac{1}{|\mathcal{S}|}\sum_{s\in\mathcal{S}}
\operatorname{Var}\!\left(\{M(\mathrm{eo},\mathrm{ec}):\mathrm{ec}\in\mathcal{C}_s^{\mathrm{eo}}\}\right).
\label{eq:app_stability}
\end{align}
These weights give each stratum equal influence. The first quantity summarizes performance; the second summarizes within-stratum sensitivity to ECs. For positive QPS observations, the geometric mean is obtained by applying the same weights to $\log_2(\mathrm{QPS})$ and exponentiating:
\begin{equation}
\operatorname{GM}_{\mathrm{QPS}}(\mathrm{eo})
=2^{\operatorname{mean}_{\log_2(\mathrm{QPS})}(\mathrm{eo})}.
\end{equation}

\paragraph{Differences under shared conditions.}
For EOs $u$ and $v$, let $\mathcal{C}_s^{uv}=\mathcal{C}_s^u\cap\mathcal{C}_s^v$ and let $\mathcal{S}_{uv}$ contain the strata with nonempty intersections. Their mean difference is
\begin{equation}
\hat\Delta_{u-v}
=\frac{1}{|\mathcal{S}_{uv}|}\sum_{s\in\mathcal{S}_{uv}}
\frac{1}{|\mathcal{C}_s^{uv}|}\sum_{\mathrm{ec}\in\mathcal{C}_s^{uv}}
\left[M(u,\mathrm{ec})-M(v,\mathrm{ec})\right].
\label{eq:app_matched_difference}
\end{equation}
The comparison uses the same ECs for both objects and retains equal stratum weights. Positive differences favor $u$ for recall and QPS; negative differences favor $u$ for latency and resource costs.

\paragraph{Exact-search reference.}
For embedding model instance $e$ and dataset instance $d$, $U(e,d,k)$ is Flat Recall@$k$ on $\operatorname{corpus}(d)$ and $\mathcal{Q}_d$, using the same similarity function and relevance judgments as the ANN evaluation. Cross-modal retrieval uses dataset-provided relevance annotations; filtered ANN uses exact neighbors within the filter-compatible candidate set. For an ANN configuration $\mathrm{eo}$ using embedding model instance $e$, let $d$ be the dataset instance specified by $\mathrm{ec}$. Define
\begin{align}
\operatorname{Gap}(\mathrm{eo},\mathrm{ec},k)
&=U(e,d,k)-R_k(\mathrm{eo},\mathrm{ec}),\\
\operatorname{Retention}(\mathrm{eo},\mathrm{ec},k)
&=\frac{R_k(\mathrm{eo},\mathrm{ec})}{U(e,d,k)},\qquad U(e,d,k)>0.
\end{align}
Flat recall measures retrieval quality under exact search. Gap and Retention describe the change relative to this reference. For a benchmark with a test5k and a full candidate pool, the two data instances are written explicitly below. For dataset instances $d_{\mathrm{test5k}}$ and $d_{\mathrm{full}}$ with the same query set and different candidate pools, Pool-Scaling Retention is
\begin{equation}
\operatorname{PSR}(e,d_{\mathrm{test5k}},d_{\mathrm{full}},k)
=\frac{U(e,d_{\mathrm{full}},k)}{U(e,d_{\mathrm{test5k}},k)},
\qquad U(e,d_{\mathrm{test5k}},k)>0.
\end{equation}

\subsubsection{Cross-Object Interactions}

\paragraph{Matched grids and weights.}
Within each index family, $\mathrm{ec}$ specifies the settings held fixed across the grid, including the dataset instance and runtime environment. The factors $E$, $I$, and $S$ represent embedding choice, construction configuration, and query-processing configuration. Their levels form a complete compatible grid
\begin{equation}
\mathcal{X}_{\mathrm{ec}}=\mathcal{E}_{\mathrm{ec}}\times\mathcal{B}_{\mathrm{ec}}\times\Gamma_{\mathrm{ec}},
\qquad
w_{\mathrm{ec}}(e,\beta,\gamma)
=\frac{1}{|\mathcal{E}_{\mathrm{ec}}|\,|\mathcal{B}_{\mathrm{ec}}|\,|\Gamma_{\mathrm{ec}}|}.
\label{eq:app_grid_weights}
\end{equation}
The dataset, candidate pool, query set, index algorithm, and runtime environment are fixed within each grid. Construction and query-processing levels are configurations of the corresponding index family, including any coupled parameter settings. The response $y_{\mathrm{ec}}(e,\beta,\gamma)$ is either Retention@10 or $\log_2(\mathrm{QPS})$, analyzed separately. Expectations below use the product weights in Equation~\ref{eq:app_grid_weights}.

\paragraph{Effect decomposition.}
Functional ANOVA decomposes the response as
\begin{equation}
\begin{aligned}
y_{\mathrm{ec}}(e,\beta,\gamma)
={}&\operatorname{mean}_{\mathrm{ec}}(y)+f_{E,\mathrm{ec}}(e)+f_{I,\mathrm{ec}}(\beta)+f_{S,\mathrm{ec}}(\gamma)\\
&+f_{EI,\mathrm{ec}}(e,\beta)+f_{ES,\mathrm{ec}}(e,\gamma)
+f_{IS,\mathrm{ec}}(\beta,\gamma)\\
&+f_{EIS,\mathrm{ec}}(e,\beta,\gamma).
\end{aligned}
\label{eq:functional_anova}
\end{equation}
Here $\operatorname{mean}_{\mathrm{ec}}(y)=\mathbb{E}[y_{\mathrm{ec}}]$ is the mean over the factor levels in the grid for $\mathrm{ec}$. For a nonempty subset $u\subseteq\{E,I,S\}$, let $x_u$ denote its factor levels and $-u$ the remaining factors. The effects are defined recursively by
\begin{equation}
f_{u,\mathrm{ec}}(x_u)
=\mathbb{E}_{-u}[y_{\mathrm{ec}}\mid x_u]-\operatorname{mean}_{\mathrm{ec}}(y)
-\sum_{\emptyset\ne v\subsetneq u}f_{v,\mathrm{ec}}(x_v).
\label{eq:app_effect_recursion}
\end{equation}
For example,
\begin{align}
f_{E,\mathrm{ec}}(e)&=\mathbb{E}_{I,S}[y_{\mathrm{ec}}\mid e]-\operatorname{mean}_{\mathrm{ec}}(y),\\
f_{ES,\mathrm{ec}}(e,\gamma)
&=\mathbb{E}_{I}[y_{\mathrm{ec}}\mid e,\gamma]-\operatorname{mean}_{\mathrm{ec}}(y)-f_{E,\mathrm{ec}}(e)-f_{S,\mathrm{ec}}(\gamma).
\end{align}
Thus, $f_{ES,\mathrm{ec}}$ measures how embedding and query processing jointly affect the response beyond their additive main effects. The three-factor term is the residual after subtracting all main and two-factor effects.

\paragraph{Variance shares.}
The effect terms have zero mean and are orthogonal under the product weights. Their variance contributions satisfy
\begin{equation}
V_{u,\mathrm{ec}}=\mathbb{E}[f_{u,\mathrm{ec}}^2],
\qquad
\operatorname{Var}[y_{\mathrm{ec}}]
=\sum_{\emptyset\ne u\subseteq\{E,I,S\}}V_{u,\mathrm{ec}}.
\end{equation}
Let $\mathcal{C}$ denote the set of evaluated EC instances. With equal weights across conditions,
\begin{equation}
\rho_u
=\frac{\sum_{\mathrm{ec}\in\mathcal{C}}V_{u,\mathrm{ec}}}
{\sum_{\mathrm{ec}\in\mathcal{C}}\operatorname{Var}[y_{\mathrm{ec}}]}.
\label{eq:interaction_share}
\end{equation}
This ratio pools variance contributions before normalization, for a positive total variance. The total interaction share is $\rho_{EI}+\rho_{ES}+\rho_{IS}+\rho_{EIS}$. The decomposition is applied separately to each build repetition, matching the per-repetition results reported in Section~\ref{subsec:exp_module_interactions}.

\paragraph{Conditional interpretation.}
Variance shares quantify interaction strength, while conditional differences describe its direction. For construction configurations $\beta_1$ and $\beta_2$, define
\begin{equation}
D_{\mathrm{ec}}(e,\gamma)
=y_{\mathrm{ec}}(e,\beta_2,\gamma)-y_{\mathrm{ec}}(e,\beta_1,\gamma).
\end{equation}
Changes in $D_{\mathrm{ec}}(e,\gamma)$ across search configurations show how query processing modifies the construction effect. A decrease in its absolute value indicates that the construction-related performance difference narrows at that search setting.

\section{Experimental Configurations}
\label{appendix:configs}

This appendix describes the configurations for the two primary retrieval scenarios in Section~\ref{sec:experiments}. 

\subsection{Runtime Environment}
\label{appendix:runtime}

The primary multimodal measurements use a server equipped with two Intel Xeon Gold 6330 CPUs and four Tesla V100-PCIE-32GB GPUs. 
These measurements use 28 CPU threads or one V100 by default. QPS variance shares in the interaction analysis are computed separately within each run. 
The VisualNews protocol for Qwen3-VL uses four separately built CAGRA graphs on V100 GPUs and merges their retrieval results.

In the filtered ANN comparisons, Flat, RangePQ, DiGRA, and ACORN-1 use one search thread. ACORN-28 uses 28 search threads. Comparisons of ACORN-28 with Flat therefore describe complete system configurations.

\subsection{Multimodal Retrieval Configurations}
\label{appendix:mm_configs}

The multimodal retrieval experiments evaluate text-to-image retrieval on Flickr30K, MSCOCO, and VisualNews. 
Flickr30K and MSCOCO include both small candidate-pool and full-pool settings, while VisualNews provides a million-scale candidate pool. 
The evaluated embedding models are CLIP, SigLIP2, and Qwen3-VL, with embedding dimensions of 512, 768, and 4096, respectively. 
Flat exact search is used as the exact-search reference under each embedding space and similarity metric. 
The evaluated ANN backends include Faiss HNSW, Faiss IVFPQ, cuVS IVFPQ, and cuVS CAGRA.

\paragraph{Measured configurations.}
Table~\ref{tab:mm_measured_grid} lists parameter levels observed in the published multimodal recall results. The interaction analysis uses the complete matched grids within these records: 330 Faiss HNSW and 165 Faiss IVFPQ configurations across five conditions, with each build repetition analyzed separately.

\begin{table*}[t]
\centering
\small
\caption{Observed parameter levels by candidate pool. The two MSCOCO pools have the same marginal levels. PQ subquantizer counts are listed in model-specific rows.}
\label{tab:mm_measured_grid}
\begin{adjustbox}{max width=\linewidth,center}
\begin{tabular}{@{}llccc>{\columncolor{white}[\tabcolsep][0pt]}c@{}}
\toprule
\textbf{Index} & \textbf{Parameter} & \textbf{Flickr30K test5k} & \textbf{Flickr30K full} & \textbf{MSCOCO (both)} & \textbf{VisualNews} \\
\midrule
Faiss HNSW &\cellcolor{vdbGreen!35} $M$ &\cellcolor{vdbGreen!35} 16, 32 &\cellcolor{vdbGreen!35} 16, 32 &\cellcolor{vdbGreen!35} 16, 32, 64 &\cellcolor{vdbGreen!35} 32, 48, 64 \\
 & \texttt{efConstruction} & 128, 256 & 128, 256 & 200, 400 & 400, 800 \\
 &\cellcolor{vdbGreen!35} \texttt{efSearch} &\cellcolor{vdbGreen!35} 32, 64, 128, 256 &\cellcolor{vdbGreen!35} 32, 64, 128, 256 &\cellcolor{vdbGreen!35} 64, 128, 256, 512 &\cellcolor{vdbGreen!35} 128, 256, 512, 1024, 2048 \\
\midrule
Faiss IVFPQ & \texttt{nlist} & 256, 512 & 256, 512, 1024 & 1024, 2048, 4096 & 4096, 8192, 16384 \\
 &\cellcolor{vdbGreen!35} \texttt{nprobe} &\cellcolor{vdbGreen!35} 4, 8, 16, 32 &\cellcolor{vdbGreen!35} 4, 8, 16, 32 &\cellcolor{vdbGreen!35} 8, 16, 32, 64, 128 &\cellcolor{vdbGreen!35} 32, 64, 128, 256, 512 \\
 & Subquantizers (CLIP) & 32, 64 & 32, 64 & 32, 64 & 64, 128 \\
 &\cellcolor{vdbGreen!35} Subquantizers (SigLIP2) &\cellcolor{vdbGreen!35} 32, 64, 96 &\cellcolor{vdbGreen!35} 32, 64, 96 &\cellcolor{vdbGreen!35} 32, 64, 96 &\cellcolor{vdbGreen!35} 64, 96, 128 \\
 & Subquantizers (Qwen3-VL) & 64, 128, 256, 512 & 64, 128, 256, 512 & 64, 128, 256, 512 & 128, 256, 512 \\
\midrule
cuVS IVFPQ &\cellcolor{vdbGreen!35} \texttt{n\_lists} &\cellcolor{vdbGreen!35} 256, 512 &\cellcolor{vdbGreen!35} 256, 512, 1024 &\cellcolor{vdbGreen!35} 1024, 2048, 4096 &\cellcolor{vdbGreen!35} 4096, 8192, 16384 \\
 & \texttt{n\_probes} & 4, 8, 16, 32 & 4, 8, 16, 32 & 8, 16, 32, 64, 128 & 32, 64, 128, 256, 512 \\
 &\cellcolor{vdbGreen!35} Subquantizers (CLIP) &\cellcolor{vdbGreen!35} 32, 64 &\cellcolor{vdbGreen!35} 32, 64 &\cellcolor{vdbGreen!35} 32, 64 &\cellcolor{vdbGreen!35} 64, 128 \\
 & Subquantizers (SigLIP2) & 32, 64, 96 & 32, 64, 96 & 32, 64, 96 & 64, 96, 128 \\
 &\cellcolor{vdbGreen!35} Subquantizers (Qwen3-VL) &\cellcolor{vdbGreen!35} 64, 128, 256, 512 &\cellcolor{vdbGreen!35} 64, 128, 256, 512 &\cellcolor{vdbGreen!35} 64, 128, 256, 512 &\cellcolor{vdbGreen!35} 128 \\
\midrule
cuVS CAGRA & Graph degree & 32, 64 & 32, 64 & 32, 64, 128 & 64, 96, 128 \\
 &\cellcolor{vdbGreen!35} Intermediate degree &\cellcolor{vdbGreen!35} 64, 128 &\cellcolor{vdbGreen!35} 64, 128 &\cellcolor{vdbGreen!35} 64, 128, 255 &\cellcolor{vdbGreen!35} 128, 255 \\
 & \texttt{itopk\_size} & 32, 64, 128, 256 & 32, 64, 128, 256 & 64, 128, 256, 512 & 128, 256, 512, 1024 \\
 &\cellcolor{vdbGreen!35} Search width &\cellcolor{vdbGreen!35} 1, 2 &\cellcolor{vdbGreen!35} 1, 2 &\cellcolor{vdbGreen!35} 1, 2, 4 &\cellcolor{vdbGreen!35} 1, 2, 4, 8 \\
\bottomrule
\end{tabular}
\end{adjustbox}
\end{table*}
Both IVFPQ implementations use 8 bits per subquantizer. CAGRA uses IVFPQ for graph construction.

\subsection{Filtered ANN Configurations}
\label{appendix:filter_configs}

The filtered ANN experiments evaluate ANN search under structured filtering conditions. 
We use 990,000 base vectors and consider two vector dimensionalities, 128 and 960, instantiated from SIFT1M- and GIST1M-derived workloads augmented with scalar attributes.
The filter ratios are \(0.01\%\), \(0.1\%\), \(1\%\), \(10\%\), and \(50\%\), covering workload regimes from highly selective filtering to loose filtering. 
The query set contains 10,000 queries.

The evaluated methods include Flat exact pre-filtering, RangePQ, DiGRA, and ACORN. 
Flat exact pre-filtering first applies the structured filter and then performs exact search within the filtered candidate set. 
\begin{table}[h]
\centering
\small
\caption{Parameter grid for RangePQ.}
\label{tab:appendix_rangepq_params}
\begin{tabular}{ll}
\toprule
\textbf{Parameter} & \textbf{Values} \\
\midrule
\rowcolor{vdbGreen!35}\(m\) & \(4, 8, 16, 32\) \\
\(\texttt{nbits}\) & \(8\) \\
\rowcolor{vdbGreen!35}\(\texttt{nlist}\) & \(512, 1024\) \\
\(\texttt{cz}\) & \(1024, 2048\) \\
\rowcolor{vdbGreen!35}\(\texttt{iter}\) & \(200\) \\
\(L\) & \(2500, 4000, 6000, 8000, 10000, 12000, 16000, 20000, 24000, 32000\) \\
\bottomrule
\end{tabular}
\end{table}

\begin{table}[h]
\centering
\small
\caption{ACORN construction grid and ACORN-28 search budgets. The ACORN-1 grid includes additional search levels ranging from 10 to 1700.}
\label{tab:appendix_acorn_params}
\begin{tabular}{ll}
\toprule
\textbf{Parameter} & \textbf{Values} \\
\midrule
\rowcolor{vdbGreen!35}\(N\) & \(990000\) \\
\(M\) & \(32, 64, 128\) \\
\rowcolor{vdbGreen!35}\(M_\beta\) & \(2M\) \\
\texttt{efConstruction} & \(100, 400\) \\
\rowcolor{vdbGreen!35}\texttt{efSearch} (ACORN-28) & \(32,48,64,80,100,128,160,200,256,320,400,512,640,800,1000,1200,1600\) \\
\(\gamma_{\mathrm{ACORN}}\), \(r=0.01\%\) & \(200, 400, 1000\) \\
\rowcolor{vdbGreen!35}\(\gamma_{\mathrm{ACORN}}\), \(r=0.1\%\) & \(200, 400\) \\
\(\gamma_{\mathrm{ACORN}}\), \(r=1\%\) & \(50, 120\) \\
\rowcolor{vdbGreen!35}\(\gamma_{\mathrm{ACORN}}\), \(r=10\%\) & \(8, 20\) \\
\(\gamma_{\mathrm{ACORN}}\), \(r=50\%\) & \(2, 6\) \\
\bottomrule
\end{tabular}
\end{table}

\begin{table}[h]
\centering
\caption{Parameter grid for DiGRA.}
\label{tab:appendix_digra_params}
\begin{tabular}{ll}
\toprule
\textbf{Parameter} & \textbf{Values} \\
\midrule
\rowcolor{vdbGreen!35}\texttt{BASE\_NUM} & \(990000\) \\
\texttt{QUERY\_NUM} & \(10000\) \\
\rowcolor{vdbGreen!35}\texttt{dim} & \(128, 960\) \\
\texttt{TOPK} & \(100\) \\
\rowcolor{vdbGreen!35}\(M\) & \(32, 64, 128\) \\
\texttt{ef\_con} & \(100, 400\) \\
\rowcolor{vdbGreen!35}\texttt{ef} & \(100, 128, 160, 200, 256, 320, 400, 512, 640, 800, 1000, 1200, 1600\) \\
\bottomrule
\end{tabular}
\end{table}

\raggedbottom
\setcounter{topnumber}{5}
\setcounter{bottomnumber}{3}
\setcounter{totalnumber}{8}
\renewcommand{\topfraction}{0.9}
\renewcommand{\bottomfraction}{0.8}
\renewcommand{\textfraction}{0.08}
\renewcommand{\floatpagefraction}{0.85}
\setlength{\floatsep}{8pt plus 1pt minus 1pt}
\setlength{\textfloatsep}{8pt plus 1pt minus 1pt}
\setlength{\intextsep}{8pt plus 1pt minus 1pt}
\makeatletter
\setlength{\@fpsep}{10pt}
\makeatother

\section{Multimodal Retrieval Results}
\label{sec:detailed_results}

\subsection{Embedding Dimensionality Controls and Significance Tests}
\label{app:embedding_controls}

We control output dimensionality to examine its contribution to differences in embedding-based retrieval quality, and assess these differences using matched-query significance tests. All measurements use Flat exact search, keeping index approximation out of the comparison.

\textbf{Matched-dimension controls.}
CLIP, SigLIP2, and Qwen3-VL produce 512-, 768-, and 4,096-dimensional embeddings, respectively. We apply two post-hoc controls to SigLIP2 and Qwen3-VL: PCA-512, fitted using corpus vectors only and applied to both corpus and query vectors, and first-512 truncation, which retains the first 512 coordinates. Both transformations are followed by $L_2$ normalization; CLIP remains unchanged at its native 512 dimensions.

The comparisons cover the Flickr30K and MSCOCO test5k and full-pool conditions, plus VisualNews. Table~\ref{tab:embedding_controls} reports the five-condition mean Flat Recall@10 under native dimensions, PCA-512, and first-512 truncation. All three settings use the same 1,446-query VisualNews subset shared by the models.

\textbf{Significance tests.}
For each model pair within each condition, we compute the difference between the two binary Flat Recall@10 outcomes for every query. Queries sharing the same relevant-image set form one cluster, giving 1,000 clusters per Flickr30K condition, 5,126 per MSCOCO condition, and 1,446 for VisualNews. The primary analysis uses a two-sided cluster-level sign-flip permutation test with 100,000 permutations and 10,000 cluster-bootstrap resamples for 95\% confidence intervals. Entire clusters are flipped or resampled together, preserving the query-weighted mean difference. We additionally use a two-sided paired $t$-test implemented as an intercept-only regression on query-level differences with CR1 cluster-robust standard errors and $G-1$ degrees of freedom, where $G$ is the number of clusters. Holm correction is applied to the 15 model-pair-by-condition comparisons separately within each dimensionality setting and test method. 

\begin{table*}[t]
\centering
\caption{Flat Recall@10 under native and matched output dimensions.}
\label{tab:embedding_controls}
\small
\begin{tabular}{lccccc}
\toprule
& \multicolumn{3}{c}{\textbf{Mean Flat Recall@10}}
& \multicolumn{2}{c}{\textbf{Largest Holm-adjusted $p$}} \\
\cmidrule(lr){2-4}\cmidrule(lr){5-6}
\textbf{Setting} & \textbf{CLIP} & \textbf{SigLIP2} & \textbf{Qwen3-VL} & \textbf{Permutation} & \textbf{Cluster-robust $t$} \\
\midrule
\rowcolor{vdbGreen!35}Native dimensions & 0.5345 & 0.2249 & 0.7235 & $1.50\times10^{-4}$ & $6.42\times10^{-10}$ \\
PCA-512 & 0.5345 & 0.1252 & 0.7213 & $1.50\times10^{-4}$ & $5.81\times10^{-10}$ \\
\rowcolor{vdbGreen!35}First-512 truncation & 0.5345 & 0.1782 & 0.7065 & $1.50\times10^{-4}$ & $5.54\times10^{-7}$ \\
\bottomrule
\end{tabular}
\end{table*}

\textbf{Results.}
Table~\ref{tab:embedding_controls} shows that model differences remain statistically significant when output dimensionality is matched. All 15 comparisons in each setting are significant under both tests after Holm correction, and all cluster-bootstrap and cluster-robust 95\% confidence intervals exclude zero. The permutation estimates reach the Monte Carlo resolution of $1/(100{,}000+1)$ before correction. The two dimensionality controls also affect the models differently: Qwen3-VL's mean Recall@10 changes from 0.7235 to 0.7213 under PCA and 0.7065 under truncation, while SigLIP2's changes from 0.2249 to 0.1252 and 0.1782. 

\subsection{Candidate-pool scaling and ANN retention}
Table~\ref{tab:pool_scaling_retention} reports the ratio of full-pool to test5k Flat Recall@10. The input recalls are given in Table~\ref{tab:flat_upper_bound}. Ratios are computed from unrounded recall values and averaged with equal dataset weights before rounding for display.

\begin{table}[t]
\centering
\caption{Pool-Scaling Retention (PSR) at cutoff 10.}
\label{tab:pool_scaling_retention}
\small
\begin{tabular}{lrrr}
\toprule
\textbf{Embedding} & \textbf{Flickr30K} & \textbf{MSCOCO} & \textbf{Mean PSR} \\
\midrule
\rowcolor{vdbGreen!35}Qwen3-VL & 77.7\% & 66.9\% & 72.3\% \\
CLIP & 54.9\% & 44.2\% & 49.6\% \\
\rowcolor{vdbGreen!35}SigLIP2 & 37.5\% & 28.9\% & 33.2\% \\
\bottomrule
\end{tabular}
\end{table}

Table~\ref{tab:ann_retention_summary} distinguishes average performance over the configuration grid from the best performance observed within each condition. Mean results first average configurations within each observed dataset-pool--embedding condition and then give each observed condition equal weight. The variance column summarizes variability across condition means. Best-configuration Retention selects the highest Retention within each condition before macro-averaging.

\begin{table}[!htbp]
\centering
\small
\renewcommand{\arraystretch}{1.15}
\caption{ANN Retention@10 and Gap@10 relative to the condition-specific Flat reference.}
\label{tab:ann_retention_summary}
\setlength{\tabcolsep}{4pt}
\begin{adjustbox}{max width=\linewidth,center}
\begin{tabular}{@{}lrrcr>{\columncolor{white}[\tabcolsep][0pt]}r@{}}
\toprule
\textbf{Backend} & \textbf{Mean Retention} & \textbf{Between-condition variance} & \textbf{95\% CI} & \textbf{Best Retention} & \textbf{Mean Gap} \\
\midrule
cuVS CAGRA & \cellcolor{vdbTeal!60}0.9959 & 0.000028 & [0.9930, 0.9983] & \cellcolor{vdbTeal!60}1.0000 & \cellcolor{vdbCoral!6}0.0012 \\
Faiss HNSW & \cellcolor{vdbTeal!59}0.9875 & 0.000201 & [0.9801, 0.9938] & \cellcolor{vdbTeal!60}0.9985 & \cellcolor{vdbCoral!7}0.0041 \\

Faiss IVFPQ & \cellcolor{vdbTeal!49}0.7932 & 0.015712 & [0.7319, 0.8538] & \cellcolor{vdbTeal!55}0.9083 & \cellcolor{vdbCoral!39}0.0751 \\
cuVS IVFPQ & \cellcolor{vdbTeal!46}0.7409 & 0.012833 & [0.6853, 0.7958] & \cellcolor{vdbTeal!53}0.8726 & \cellcolor{vdbCoral!52}0.1015 \\
\bottomrule
\end{tabular}
\end{adjustbox}
\end{table}

\subsection{Throughput under matched quality targets}
Table~\ref{tab:app_flat_qps} reports Flat throughput by condition for CLIP (512 dimensions), SigLIP2 (768 dimensions), and Qwen3-VL (4096 dimensions). Repeated Flat QPS observations within the same condition are averaged. Ratios are computed from unrounded values.

\begin{table}[!htbp]
\centering
\small
\renewcommand{\arraystretch}{1.15}
\caption{Flat QPS for the three embedding models.}
\label{tab:app_flat_qps}
\setlength{\tabcolsep}{4pt}
\begin{adjustbox}{max width=\linewidth,center}
\begin{tabular}{@{}lrrrr@{}}
\toprule
\textbf{Condition} & \textbf{CLIP} & \textbf{SigLIP2} & \textbf{Qwen3-VL} & \textbf{CLIP/Qwen3-VL} \\
\midrule
Flickr30K test5k & \cellcolor{vdbTeal!60}9,305.00 & \cellcolor{vdbTeal!54}7,718.89 & \cellcolor{vdbTeal!5}1,662.53 & $5.60\times$ \\
Flickr30K full-pool & \cellcolor{vdbTeal!60}568.94 & \cellcolor{vdbTeal!50}381.01 & \cellcolor{vdbTeal!5}61.64 & $9.23\times$ \\
MSCOCO test5k & \cellcolor{vdbTeal!60}3,491.97 & \cellcolor{vdbTeal!51}2,373.31 & \cellcolor{vdbTeal!5}348.16 & $10.03\times$ \\
MSCOCO full-pool & \cellcolor{vdbTeal!60}207.18 & \cellcolor{vdbTeal!50}139.68 & \cellcolor{vdbTeal!5}24.94 & $8.31\times$ \\
VisualNews & \cellcolor{vdbTeal!60}13.81 & \cellcolor{vdbTeal!50}9.22 & \cellcolor{vdbTeal!5}1.61 & $8.55\times$ \\
\bottomrule
\end{tabular}
\end{adjustbox}
\end{table}

Table~\ref{tab:app_visualnews_points} lists the highest-throughput measured configuration of each backend on VisualNews with CLIP, subject to $\mathrm{Retention}@10 \geq 0.99$. Here $\mathrm{efC}$ and $\mathrm{efS}$ denote \texttt{efConstruction} and \texttt{efSearch}. The throughput measurements use a separate VisualNews candidate pool, with Retention computed against its corresponding Flat reference.

\begin{table}[!htbp]
\centering
\small
\renewcommand{\arraystretch}{1.15}
\caption{Highest-throughput configurations satisfying $\mathrm{Retention}@10 \geq 0.99$ on VisualNews with CLIP.}
\label{tab:app_visualnews_points}
\setlength{\tabcolsep}{3pt}
\begin{tabularx}{\linewidth}{@{}lrrrl>{\raggedright\arraybackslash}X@{}}
\toprule
\textbf{Backend} & \textbf{Recall@10} & \textbf{Retention@10} & \textbf{QPS} & \textbf{Resources} & \textbf{Parameters} \\
\midrule
Flat & 0.29895 & 1.00000 & \cellcolor{vdbTeal!10}13.81 & CPU, 28 threads & Exact search \\
Faiss HNSW & 0.29605 & 0.99030 & \cellcolor{vdbTeal!42}1,898.56 & CPU, 28 threads & $M=32$, $\mathrm{efC}=400$, $\mathrm{efS}=512$ \\
cuVS CAGRA & 0.29680 & 0.99281 & \cellcolor{vdbTeal!60}29,433.58 & One V100 & Graph degree $=64$, intermediate degree $=128$, itopk size $=256$, search width $=2$ \\
\bottomrule
\end{tabularx}
\end{table}

\section{Filtered ANN Results}
\subsection{Real workloads and random seed settings}
\label{app:filtered_parameterization}

\textbf{Real workloads.}
We evaluate ACORN on CC-News (620,643 vectors) with 3-day and 30-day timestamp windows and on App Reviews (277,936 vectors) with conjunctions of three attribute-range predicates. Each workload uses 10,000 queries, 384-dimensional vectors, and exact filtered-neighbor IDs. We evaluate 156 configurations per workload using $M\in\{32,64,128\}$, $M_\beta=2M$, $\gamma_{\mathrm{ACORN}}\in\{2,6\}$, \texttt{efConstruction}$\in\{100,400\}$, and 13 search budgets from 100 to 1600, with 28 search threads.

\begin{table*}[t]
\centering
\caption{ACORN construction-parameter audits on real workloads. Drops are absolute decreases in Recall@100 under matched ECs. The $M$ comparisons jointly change $M_\beta=2M$.}
\label{tab:real_filtered_parameter_audit}
\small
\begin{tabular}{llrrrr}
\toprule
\textbf{Workload} & \textbf{Parameter} & \textbf{Comparisons} & \textbf{Drops $>0$} & \textbf{Drops $>10^{-4}$} & \textbf{Max drop} \\
\midrule
CC-News, 3-day &\cellcolor{vdbGreen!35} $M$ &\cellcolor{vdbGreen!35} 104 &\cellcolor{vdbGreen!35} 8 &\cellcolor{vdbGreen!35} 8 &\cellcolor{vdbGreen!35} 0.0134 \\
 & \texttt{efConstruction} & 78 & 39 & 39 & 0.3107 \\
CC-News, 30-day &\cellcolor{vdbGreen!35} $M$ &\cellcolor{vdbGreen!35} 104 &\cellcolor{vdbGreen!35} 15 &\cellcolor{vdbGreen!35} 14 &\cellcolor{vdbGreen!35} 0.0115 \\
 & \texttt{efConstruction} & 78 & 39 & 39 & 0.3380 \\
App Reviews &\cellcolor{vdbGreen!35} $M$ &\cellcolor{vdbGreen!35} 104 &\cellcolor{vdbGreen!35} 24 &\cellcolor{vdbGreen!35} 24 &\cellcolor{vdbGreen!35} 0.0887 \\
 & \texttt{efConstruction} & 78 & 35 & 35 & 0.3061 \\
\bottomrule
\end{tabular}
\end{table*}

Increasing \texttt{efConstruction} from 100 to 400 lowers Recall@100 in 39 of 78 matched comparisons for each CC-News window and in 35 of 78 on App Reviews (Table~\ref{tab:real_filtered_parameter_audit}).

\textbf{Random seed settings.}
The ACORN graph construction seed for both CC-News windows was 12345. The App Reviews driver initialized the C random number generator with \texttt{srand(0)}. Each real-workload configuration used one index build. For the SIFT128 graph analysis, we built each of the four ACORN configurations twice using the same graph construction seed (12345) and 112 construction threads.

\subsection{Throughput and construction cost}

Table~\ref{tab:app_filtered_qps} reports the highest QPS at $\mathrm{Recall}@100 \geq 0.95$ for each dimensionality, filter ratio, and method; a dash indicates no qualifying configuration. ACORN-28 uses 28 search threads; others use one. Speedup is the ACORN-28/Flat QPS ratio. On 128-dimensional SIFT, ACORN-28 achieves a median $14.243\times$ QPS speedup over ACORN-1 across 1,122 pairs matched on filter ratio, construction parameters, and search budget.

\begin{table}[!htbp]
\centering
\small
\renewcommand{\arraystretch}{1.15}
\caption{Highest measured QPS subject to $\mathrm{Recall}@100 \geq 0.95$.}
\label{tab:app_filtered_qps}
\setlength{\tabcolsep}{3pt}
\begin{adjustbox}{max width=\linewidth,center}
\begin{tabular}{@{}>{\columncolor{white}[0pt][\tabcolsep]}rrrrrrr>{\columncolor{white}[\tabcolsep][0pt]}r@{}}
\toprule
\textbf{dim} & \textbf{Filter ratio} & \textbf{Flat} & \textbf{RangePQ} & \textbf{DiGRA} & \textbf{ACORN-1} & \textbf{ACORN-28} & \textbf{Speedup} \\
\midrule
\rowcolor{vdbGreen!35}128 & 0.01\% & 8,949.74 & 4,483.70 & 4,238.16 & --- & --- & --- \\
128 & 0.1\% & 3,913.89 & --- & 2,609.59 & 77.80 & 1,416.40 & $0.4\times$ \\
\rowcolor{vdbGreen!35}128 & 1\% & 537.43 & --- & 1,789.91 & 431.20 & 5,821.10 & $10.8\times$ \\
128 & 10\% & 48.07 & --- & 1,592.11 & 687.50 & 9,678.30 & $201.3\times$ \\
\rowcolor{vdbGreen!35}128 & 50\% & 9.87 & --- & 1,348.07 & 610.40 & 10,377.60 & $1,051.7\times$ \\
\midrule
960 & 0.01\% & 1,158.62 & --- & 3,843.01 & --- & --- & --- \\
\rowcolor{vdbGreen!35}960 & 0.1\% & 471.99 & --- & 2,246.98 & --- & --- & --- \\
960 & 1\% & 32.41 & --- & 1,250.22 & 35.70 & 3,157.80 & $97.4\times$ \\
\rowcolor{vdbGreen!35}960 & 10\% & 2.83 & --- & 1,200.50 & 23.20 & 1,767.30 & $624.5\times$ \\
960 & 50\% & 0.53 & --- & 904.78 & 15.20 & 977.80 & $1,842.9\times$ \\
\bottomrule
\end{tabular}
\end{adjustbox}
\end{table}

%To assess parallel execution effects, we additionally compare ACORN-28 with ACORN-1 under matched configurations. On 128-dimensional SIFT, 1,122 pairs share the same filter ratio, construction parameters, and search budget. Their QPS ratios have a median of $14.243\times$ and an interquartile range of $[9.87\times,16.83\times]$. Individual pairs and statistics by filter ratio are available in the GitHub repository.

Beyond query throughput, Table~\ref{tab:app_build_cost} reports construction time and peak RSS across construction configurations. DiGRA requires substantially longer construction times than RangePQ and ACORN-1. ACORN-1 has the lowest median construction time, while RangePQ has the lowest median peak RSS among the reported methods at both dimensionalities.

\begin{table}[!htbp]
\centering
\small
\renewcommand{\arraystretch}{1.15}
\caption{Construction time (s) and peak RSS (MB) by method and dimensionality.}
\label{tab:app_build_cost}
\setlength{\tabcolsep}{3pt}
\begin{adjustbox}{max width=\linewidth,center}
\begin{tabular}{@{}>{\columncolor{white}[0pt][\tabcolsep]}rlrrrr>{\columncolor{white}[\tabcolsep][0pt]}r@{}}
\toprule
\textbf{dim} & \textbf{Method} & \textbf{Configs.} & \textbf{Median time} & \textbf{Time range} & \textbf{Median RSS} & \textbf{RSS range} \\
\midrule
\rowcolor{vdbGreen!35}128 & RangePQ & 80 & 21.49 & 5.87--111.61 & 1,136.3 & 1,085.8--1,594.0 \\
128 & DiGRA & 30 & 6,162.48 & 2,291.21--14,852.00 & 6,106.1 & 3,689.1--10,940.3 \\
\rowcolor{vdbGreen!35}128 & ACORN-1 & 66 & 3.71 & 1.44--7.90 & 4,426.3 & 1,877.0--12,004.1 \\
\midrule
\rowcolor{vdbGreen!35}960 & RangePQ & 80 & 27.49 & 8.73--125.88 & 8,894.0 & 7,143.1--11,496.6 \\
960 & DiGRA & 30 & 7,772.06 & 2,879.72--19,788.00 & 9,280.2 & 6,863.1--14,114.3 \\
\rowcolor{vdbGreen!35}960 & ACORN-1 & 60 & 10.75 & 6.03--36.55 & 10,523.4 & 8,129.5--13,759.5 \\
\bottomrule
\end{tabular}
\end{adjustbox}
\end{table}

\section{Additional Retrieval Scenarios}
\label{app:additional_scenarios}

We extend VecDBLens to Chinese text retrieval and multi-vector retrieval. Task quality is measured using dataset relevance judgments. ANN@100 measures recovery of the corresponding exact top-100 results.

\subsection{Chinese Text Retrieval}
\label{app:miracl_zh}

We evaluate Qwen3-VL (4096 dimensions) and SigLIP2 (768 dimensions) on MIRACL Chinese~\cite{zhang2023miracl}, using 4,934,368 passages and 393 development queries. Embeddings are normalized. Each model is evaluated with Flat, five HNSW search budgets ($\texttt{efSearch}=128,256,512,1024,2048$), and 45 IVFPQ configurations. The IVFPQ grid combines $\texttt{nlist}\in\{4096,8192,16384\}$, $\texttt{nprobe}\in\{32,64,128,256,512\}$, and three subquantizer counts: $\{64,96,128\}$ for SigLIP2 and $\{128,256,512\}$ for Qwen3-VL, with 8 bits per subquantizer. QPS is the median of three full-query-batch measurements using 28 CPU threads, excluding encoding.

Table~\ref{tab:supp_miracl_summary} separates model-dependent task quality from ANN approximation. HNSW preserves each model's task metrics at the displayed precision across search budgets, while IVFPQ introduces additional degradation. High exact-neighbor recovery therefore does not compensate for low representation quality. The model comparison includes their encoding settings: SigLIP2's 64-token limit truncates approximately 56.55\% of the corpus.

\begin{table}[!htbp]
\centering
\small
\setlength{\tabcolsep}{4pt}
\caption{MIRACL Chinese results. Ranges give the minimum and maximum of each metric across all tested configurations; extrema need not occur at the same configuration. }
\label{tab:supp_miracl_summary}
\begin{adjustbox}{max width=\linewidth,center}
\begin{tabular}{@{}>{\columncolor{white}[0pt][\tabcolsep]}llrrr>{\columncolor{white}[\tabcolsep][0pt]}r@{}}
\toprule
\textbf{Model} & \textbf{Method} & \textbf{nDCG@10} & \textbf{Recall@100} & \textbf{ANN@100} & \textbf{QPS} \\
\midrule
\rowcolor{vdbGreen!35}SigLIP2 & Flat & 0.0057 & 0.0217 & 1.0000 & 247.2 \\
SigLIP2 & HNSW & 0.0057 & 0.0217 & 0.9468--0.9936 & 1060.8--16642.2 \\
\rowcolor{vdbGreen!35}SigLIP2 & IVFPQ & 0.0006--0.0035 & 0.0061--0.0138 & 0.3393--0.5678 & 167.8--6190.4 \\
\midrule
Qwen3-VL & Flat & 0.5973 & 0.9319 & 1.0000 & 44.4 \\
\rowcolor{vdbGreen!35}Qwen3-VL & HNSW & 0.5973 & 0.9319 & 0.9911--0.9992 & 46.2--2691.1 \\
Qwen3-VL & IVFPQ & 0.4819--0.5782 & 0.7703--0.9107 & 0.5438--0.7892 & 58.2--1494.6 \\
\bottomrule
\end{tabular}
\end{adjustbox}
\end{table}

\begin{table}[htbp]
\centering
\small
\setlength{\tabcolsep}{4pt}
\caption{Multi-vector retrieval at the two endpoints of the evaluated candidate-budget range.}
\label{tab:supp_multivector}
\begin{adjustbox}{max width=\linewidth,center}
\begin{tabular}{@{}llrrr>{\columncolor{white}[\tabcolsep][0pt]}r@{}}
\toprule
\textbf{Dataset} & \textbf{Configuration} & \textbf{nDCG@10} & \textbf{Recall@100} & \textbf{ANN@100} & \textbf{Latency (ms)} \\
\midrule
SciFact & Exact MaxSim & \cellcolor{vdbTeal!41}0.6464 & \cellcolor{vdbTeal!54}0.8942 & \cellcolor{vdbTeal!60}1.0000 & \cellcolor{vdbTeal!16}9.89 \\
SciFact & PLAID, $K=100$ & \cellcolor{vdbTeal!41}0.6465 & \cellcolor{vdbTeal!54}0.8909 & \cellcolor{vdbTeal!54}0.8884 & \cellcolor{vdbTeal!60}6.69 \\
SciFact & PLAID, $K=1000$ & \cellcolor{vdbTeal!41}0.6464 & \cellcolor{vdbTeal!54}0.8942 & \cellcolor{vdbTeal!60}0.9953 & \cellcolor{vdbTeal!5}10.89 \\
\midrule
FiQA & Exact MaxSim & \cellcolor{vdbTeal!24}0.3472 & \cellcolor{vdbTeal!40}0.6277 & \cellcolor{vdbTeal!60}1.0000 & \cellcolor{vdbTeal!5}109.27 \\
FiQA & PLAID, $K=100$ & \cellcolor{vdbTeal!24}0.3457 & \cellcolor{vdbTeal!39}0.6201 & \cellcolor{vdbTeal!54}0.8825 & \cellcolor{vdbTeal!60}7.20 \\
FiQA & PLAID, $K=1000$ & \cellcolor{vdbTeal!24}0.3456 & \cellcolor{vdbTeal!39}0.6235 & \cellcolor{vdbTeal!58}0.9713 & \cellcolor{vdbTeal!50}11.65 \\
\bottomrule
\end{tabular}
\end{adjustbox}
\end{table}

\subsection{Multi-Vector Retrieval}
\label{app:multivector}

We evaluate ColBERTv2~\cite{santhanam2022colbertv2} on BEIR SciFact and FiQA~\cite{thakur2021beir, wadden2020fact, maia201818}, with 5,183/57,638 documents and 300/648 test queries, respectively. Token vectors have 128 dimensions, with query/document lengths of 32/180 and the model's masking and normalization. Exact search applies MaxSim~\cite{khattab2020colbert} to uncompressed token representations over the full corpus. PLAID~\cite{santhanam2022plaid} generates candidates with 2-bit compression and reranks them using the same uncompressed representations, with budgets $K\in\{100,250,500,1000\}$. Measurements use one V100 GPU and 16 CPU threads; latency excludes encoding.

Table~\ref{tab:supp_multivector} reports the exact reference and the smallest and largest evaluated candidate budgets. On FiQA, $K=100$ reduces mean latency from 109.27 to 7.20\,ms, with an nDCG@10 difference of $-0.001452$ (pointwise 95\% query-bootstrap interval: $[-0.004604,0.000182]$). Larger budgets improve exact-neighbor recovery without consistent gains in task quality. On SciFact, $K=1000$ is slower than exact search, showing how corpus size and search budget affect the quality--cost trade-off.

\clearpage
\newpage
\section*{NeurIPS Paper Checklist}

\begin{enumerate}

\item {\bf Claims}
    \item[] Question: Do the main claims made in the abstract and introduction accurately reflect the paper's contributions and scope?
    \item[] Answer: \answerYes{} % Replace by \answerYes{}, \answerNo{}, or \answerNA{}.
    \item[] Justification: The abstract and introduction describe VecDBLens as a framework for individual object attribution and cross-object interaction analysis. Section~\ref{sec:experiments} evaluates the framework in two primary retrieval scenarios, with additional scenarios reported in Appendix~\ref{app:additional_scenarios}.

\item {\bf Limitations}
    \item[] Question: Does the paper discuss the limitations of the work performed by the authors?
    \item[] Answer: \answerYes{} % Replace by \answerYes{}, \answerNo{}, or \answerNA{}.
    \item[] Justification: We explicitly discuss the limitations of our framework, including its current reliance on system-level metrics rather than fine-grained semantic diagnostics and the specific scope of the evaluated workloads.

\item {\bf Theory assumptions and proofs}
    \item[] Question: For each theoretical result, does the paper provide the full set of assumptions and a complete (and correct) proof?
    \item[] Answer: \answerNA{} % Replace by \answerYes{}, \answerNo{}, or \answerNA{}.
    \item[] Justification: The paper does not present new theorems or formal proof-based claims. Its mathematical content consists of framework definitions and metrics for individual object attribution and cross-object interaction analysis, with notation and formal definitions provided in the main text and appendix.

\item {\bf Experimental result reproducibility}
    \item[] Question: Does the paper fully disclose all the information needed to reproduce the main experimental results of the paper to the extent that it affects the main claims and/or conclusions of the paper (regardless of whether the code and data are provided or not)?
    \item[] Answer: \answerYes{} % Replace by \answerYes{}, \answerNo{}, or \answerNA{}.
    \item[] Justification:  The paper describes the evaluated datasets, embedding models, index backends, filtered ANN methods, parameter grids, evaluation metrics, statistical protocol, hardware, and runtime settings. The appendix further provides full experimental configurations and detailed numerical results supporting the main claims.

\item {\bf Open access to data and code}
    \item[] Question: Does the paper provide open access to the data and code, with sufficient instructions to faithfully reproduce the main experimental results, as described in supplemental material?
    \item[] Answer: \answerYes{} % Replace by \answerYes{}, \answerNo{}, or \answerNA{}.
    \item[] Justification: The paper provides a publicly accessible code repository link. The experiments are based on public datasets, publicly available embedding models, and open-source or publicly available vector search systems; the appendix and repository provide the configurations needed to reproduce the reported results.

\item {\bf Experimental setting/details}
    \item[] Question: Does the paper specify all the training and test details (e.g., data splits, hyperparameters, how they were chosen, type of optimizer) necessary to understand the results?
    \item[] Answer: \answerYes{} 
    \item[] Justification: The paper specifies the retrieval datasets, candidate-pool settings, embedding dimensions, index algorithms, construction and search parameter grids, filter ratios, thread settings, GPU settings, and evaluation metrics. Since the paper evaluates existing embedding models and ANN systems rather than training new models, optimizer details are not applicable.

\item {\bf Experiment statistical significance}
    \item[] Question: Does the paper report error bars suitably and correctly defined or other appropriate information about the statistical significance of the experiments?
    \item[] Answer: \answerYes{} % Replace by \answerYes{}, \answerNo{}, or \answerNA{}.
    \item[] Justification: The paper reports mean performance, variance, and 95\% bootstrap confidence intervals. Differences between embedding models are assessed using cluster-aware significance tests with Holm correction for multiple comparisons, as detailed in Appendix~\ref{app:embedding_controls}.

\item {\bf Experiments compute resources}
    \item[] Question: For each experiment, does the paper provide sufficient information on the computer resources (type of compute workers, memory, time of execution) needed to reproduce the experiments?
    \item[] Answer: \answerYes{}% Replace by \answerYes{}, \answerNo{}, or \answerNA{}.
    \item[] Justification: The paper describes a shared runtime environment used across all experiments, including CPU/GPU hardware, default thread settings, and GPU usage. It also specifies experiment-specific execution details when they differ from the default setup, such as four-GPU graph construction for high-dimensional Qwen3-VL embeddings on VisualNews and separate ACORN-1/ACORN-28 settings for filtered ANN search.

\item {\bf Code of ethics}
    \item[] Question: Does the research conducted in the paper conform, in every respect, with the NeurIPS Code of Ethics \url{https://neurips.cc/public/EthicsGuidelines}?
    \item[] Answer: \answerYes{} % Replace by \answerYes{}, \answerNo{}, or \answerNA{}.
    \item[] Justification:The research is an empirical evaluation study of vector database retrieval systems using public datasets, public or cited embedding models, and open-source or publicly available vector search systems. To the best of our knowledge, the work conforms to the NeurIPS Code of Ethics.

\item {\bf Broader impacts}
    \item[] Question: Does the paper discuss both potential positive societal impacts and negative societal impacts of the work performed?
    \item[] Answer: \answerYes{} % Replace by \answerYes{}, \answerNo{}, or \answerNA{}.
    \item[] Justification:  The paper discusses how more transparent vector database evaluation can help practitioners identify true retrieval bottlenecks and avoid misleading performance conclusions.

\item {\bf Safeguards}
    \item[] Question: Does the paper describe safeguards that have been put in place for responsible release of data or models that have a high risk for misuse (e.g., pre-trained language models, image generators, or scraped datasets)?
    \item[] Answer: \answerNA{} % Replace by \answerYes{}, \answerNo{}, or \answerNA{}.
    \item[] Justification: The paper does not release high-risk pretrained models, image generators, or newly scraped datasets. It evaluates existing retrieval datasets, embedding models, and vector search systems.

\item {\bf Licenses for existing assets}
    \item[] Question: Are the creators or original owners of assets (e.g., code, data, models), used in the paper, properly credited and are the license and terms of use explicitly mentioned and properly respected?
    \item[] Answer: \answerYes{} % Replace by \answerYes{}, \answerNo{}, or \answerNA{}.
    \item[] Justification:The paper cites the datasets, embedding models, ANN libraries, and filtered ANN systems used in the experiments, including their original sources where available. The experiments are conducted using existing public or publicly available assets under their respective access conditions.

\item {\bf New assets}
    \item[] Question: Are new assets introduced in the paper well documented and is the documentation provided alongside the assets?
    \item[] Answer: \answerNA{} % Replace by \answerYes{}, \answerNo{}, or \answerNA{}.
    \item[] Justification: The paper does not introduce a new dataset or pretrained model as a released asset. The contribution is a modular evaluation and attribution framework together with empirical findings based on existing assets.

\item {\bf Crowdsourcing and research with human subjects}
    \item[] Question: For crowdsourcing experiments and research with human subjects, does the paper include the full text of instructions given to participants and screenshots, if applicable, as well as details about compensation (if any)? 
    \item[] Answer: \answerNA{} % Replace by \answerYes{}, \answerNo{}, or \answerNA{}.
    \item[] Justification:The paper does not involve crowdsourcing experiments or research with human subjects.

\item {\bf Institutional review board (IRB) approvals or equivalent for research with human subjects}
    \item[] Question: Does the paper describe potential risks incurred by study participants, whether such risks were disclosed to the subjects, and whether Institutional Review Board (IRB) approvals (or an equivalent approval/review based on the requirements of your country or institution) were obtained?
    \item[] Answer: \answerNA{} % Replace by \answerYes{}, \answerNo{}, or \answerNA{}.
    \item[] Justification: The paper does not involve human subjects, user studies, or crowdsourcing experiments, so IRB approval is not applicable.

\item {\bf Declaration of LLM usage}
    \item[] Question: Does the paper describe the usage of LLMs if it is an important, original, or non-standard component of the core methods in this research? Note that if the LLM is used only for writing, editing, or formatting purposes and does \emph{not} impact the core methodology, scientific rigor, or originality of the research, declaration is not required.
    %this research? 
    \item[] Answer: \answerNA{} % Replace by \answerYes{}, \answerNo{}, or \answerNA{}.
    \item[] Justification:AI/LLM tools were used only for writing assistance, including language polishing, editing, and formatting. They were not used as an important, original, or non-standard component of the core method, and did not affect the experimental design, implementation, analysis, scientific rigor, or originality of the research.

\end{enumerate}

\end{document}